\documentclass[conference]{IEEEtran}
\IEEEoverridecommandlockouts

\usepackage{graphicx}
\usepackage{hyperref}
\usepackage{xspace}
\usepackage{amsmath}
\usepackage{cleveref}
\usepackage{listings}
\usepackage{color}
\usepackage{comment}
\usepackage{url}
\usepackage{subfig}
\usepackage{times}
\begin{document}

\title{PushdownDB: Accelerating a DBMS \\using S3 Computation}

\author{
\IEEEauthorblockN{
	Xiangyao Yu\IEEEauthorrefmark{1}, 
	Matt Youill\IEEEauthorrefmark{3}, 
	Matthew Woicik\IEEEauthorrefmark{2},
	Abdurrahman Ghanem\IEEEauthorrefmark{4}, \\
	Marco Serafini\IEEEauthorrefmark{5},
	Ashraf Aboulnaga\IEEEauthorrefmark{4}, 
	Michael Stonebraker\IEEEauthorrefmark{2}
}
\IEEEauthorblockA{
\small
	\IEEEauthorrefmark{1}University of Wisconsin-Madison
	\IEEEauthorrefmark{2}Massachusetts Institute of Technology \\
	\IEEEauthorrefmark{3}Burnian 
	\IEEEauthorrefmark{4}Qatar Computing Research Institute
	\IEEEauthorrefmark{5}University of Massachusetts Amherst
}
\normalsize
Email: 
	yxy@cs.wisc.edu, 
	matt.youill@burnian.com, 
	mwoicik@mit.edu,
	abghanem@hbku.edu.qa, \\
	marco@cs.umass.edu,
	aaboulnaga@hbku.edu.qa,
	stonebraker@csail.mit.edu
}

\date{}
\maketitle

\newcommand\name{\textit{PushdownDB}\xspace}
\newcommand{\todo}[1]{\textcolor{red}{#1}}
\newcommand{\spara}[1]{\vspace{1mm}\noindent\textbf{#1.}}

\begin{abstract}

This paper studies the effectiveness of pushing parts of DBMS analytics queries 
into the Simple Storage Service (S3) engine of Amazon Web Services 
(AWS), using a recently released capability called S3 Select.  We show 
that some DBMS primitives (filter, projection, aggregation) can always 
be cost-effectively moved into S3.  Other more complex operations 
(join, top-K, group-by) require reimplementation to take advantage of 
S3 Select and are often candidates for pushdown.  We demonstrate these 
capabilities through experimentation using a new DBMS that we 
developed, \name.  Experimentation with a collection of queries 
including TPC-H queries shows that \name is on average 30\% cheaper 
and 6.7$\times$ faster than a baseline that does not use S3 Select.  
\end{abstract}


\section{Introduction} \label{sec:introduction}

Clouds offer cheaper and more flexible computing than ``on-prem''. Not only can one add resources on the fly, the large cloud vendors have major economies of scale relative to ``on-prem'' deployment. Modern clouds employ an architecture where the computation and storage are disaggregated --- the two components are independently managed and connected using a network. Such an architecture allows for independent scaling of computation and storage, which simplifies the management of storage and reduces its cost. A number of data warehousing systems have been built to analyze data on disaggregated cloud storage, including Presto~\cite{presto}, Snowflake~\cite{snowflake}, Redshift Spectrum~\cite{spectrum}, among others.

In a disaggregated architecture, the network that connects the computation and storage layers can be a major performance bottleneck. The internal bandwidth of the storage devices within a storage server is much higher than the external network bandwidth commonly offered by cloud storage. As a result, a database running on a disaggregated architecture may underperform a database on a conventional shared-nothing architecture, where the storage devices are attached to the compute servers themselves~\cite{clouddb}.

Two intuitive solutions exist to mitigate the network bottleneck: \textit{caching} and \textit{computation pushdown}. With caching, a compute server loads data from the remote storage once, caches it in main memory or local storage, and reuses it across multiple queries, thereby amortizing the network transfer cost. Caching has been implemented in Snowflake~\cite{snowflake} and the Redshift~\cite{gupta2015amazon} layer of Redshift Spectrum~\cite{spectrum}. With computation pushdown, a database management system (DBMS) pushes its functionality as close to storage as possible.
A pioneering paper by Hagmann~\cite{hagmann1986performance} studied the division of SQL code between the storage layer and the application layer and concluded that performance was optimized if all code was moved into the storage layer. Moreover, one of the design tenets of the Britton-Lee IDM 500~\cite{ubell1985intelligent}, the Oracle Exadata server~\cite{exadata}, and the IBM Netezza machine~\cite{netezza} was to push computation into specialized processors that are closer to storage.

Recently, Amazon Web Services (AWS) introduced a feature called ``S3 Select'', through which limited computation can be pushed onto their shared cloud storage service called S3~\cite{s3select}. This provides an opportunity to revisit the question of how to divide query processing tasks between S3 storage nodes and normal computation nodes.
The question is nontrivial as the limited computational interface of S3 Select allows only certain simple query operators to be pushed into the storage layer, namely selections, projections, and simple aggregations. Other operators require new implementations to take advantage of S3 Select. Moreover, S3 Select pricing can be more expensive than computing on normal EC2 nodes.

In this paper, we set our goal to understand the performance of computation pushdown when running queries in a cloud setting with disaggregated storage.
Specifically, we consider filter (with and without indexing), join, group-by, and top-K as candidates. We implement these operators to take advantage of computation pushdown through S3 Select and study their cost and performance. We show dramatic performance improvement and cost reduction, even with the relatively high cost of S3 Select.
In addition, we analyze queries from the TPC-H benchmark and show similar benefits of performance and cost. We also point out the limitations of the current S3 Select service and provide several suggestions based on the lessons we learned from this project.
To the best of our knowledge, this is the \textit{first extensive study of pushdown computing for database operators in a disaggregated architecture.}

For the rest of this paper, Section~\ref{sec:experimental} describes the cloud environment of our evaluation.
Section~\ref{sec:testbed} describes the \name database we implemented.
Then Sections~\ref{sec:filter}--\ref{sec:topk} describe how filter, join, group-by, and top-K can leverage S3 Select, and evaluates the performance using micro benchmarks.
Section~\ref{sec:tpch} shows evaluation on the TPC-H benchmark suite.
Section~\ref{sec:parquet} evaluates the Parquet data format.
Section~\ref{sec:discussion} discusses ways to improve the current S3 Select interface.
Finally, Section~\ref{sec:related} describes related work and Section~\ref{sec:conclusion} concludes the paper.


\section{Data Management in the Cloud} 
\label{sec:experimental}

Cloud providers such as AWS offer a wide variety of computing services, and renting nodes is a basic one. In AWS, this service is called \textit{Elastic Compute Cloud} (EC2).
EC2 computing nodes (called \textit{instances}) come in different configurations and can have locally-attached storage.

In the context of a DBMS, EC2 instances are used to execute SQL queries. 
AWS offers \textit{Simple Storage Service} (S3)~\cite{s3}, a highly available object store. 
S3 provides virtually infinite storage capacity for regular users with relatively low cost, and is supported by most popular cloud databases, including Presto~\cite{presto}, Hive~\cite{hive}, Spark SQL~\cite{sparksql}, Redshift Spectrum~\cite{spectrum}, and Snowflake~\cite{snowflake}.
The storage nodes in S3 are separate from compute nodes. Hence, a DBMS uses S3 as a storage system and transfers needed data over a network for query processing.

S3 is a popular storage choice for cloud databases, since S3 storage is much cheaper than locally-attached and/or block-based alternatives, e.g., \textit{Elastic Block Store} (EBS). In addition, S3 data can be shared across multiple computing instances. 

\subsection{S3 Select} \label{ssec:select}

To reduce network traffic and the associated processing on compute nodes, AWS released a new service called \textit{S3 Select}~\cite{s3select} in 2018 to push limited computation to the storage nodes.  
Normal S3 supports put/get operators that write/read a whole object or part of it (based on byte offsets). 
S3 Select adds support for a limited set of SQL queries. At the current time, S3 Select supports only selection, projection, and aggregation without group-by for tables using the CSV or Parquet~\cite{parquet} format.

We show examples of the SQL queries supported by S3 Select in the subsequent sections. 
S3 Select implements these operators by scanning the rows in the table and returning qualifying rows to the compute node. More sophisticated operations such as join, group by, and top-K are not supported by S3 Select and need to be executed at a compute node. Redesigning these more complex query operators to use S3 Select is challenging.  For example, supporting a join operator will require data shuffling among storage nodes. In this paper, we study how these advanced operators can be broken down into simpler ones to leverage S3 Select. We propose and evaluate several implementations of these more advanced operators and show that they can often be made faster and cheaper than loading all data into EC2 compute nodes. 

\subsection{Computing Query Cost} 
\label{ssec:cost}

The dollar cost of queries is a crucial factor, since it is one of the main reasons to migrate an application from ``on-prem'' to the cloud. 
For the same AWS service, cost varies based on the region where the user’s data and computation are located. 
We limit our cost calculation to US East (N. Virginia) pricing.
In this section, we discuss the costs associated with the services we use in our experiments: storage, data access, data transfer, network requests, and computation on EC2 instances.

\spara{Storage cost}
S3 storage cost is charged monthly based on the amount of space used. For example, S3 standard storage costs about \$0.022/GB/month. Although other AWS storage services may offer better IO performance, they are also more expensive than S3. Since the storage cost only depends on data size and not on frequency of access, we exclude it when calculating query cost in this paper.

\spara{Data transfer cost}
AWS S3 users are charged for only the outgoing traffic and the price is based on the destination of the data. 
When S3 Select is \textbf{not} used, this price ranges from free (transferring data within the same region) to \$0.09/GB (transferring data out of AWS). 
Servers in our experiments are within the same region as the S3 data. Therefore, we do not pay any data transfer cost.

\spara{S3 Select cost}
S3 Select introduces a new cost component that is based on the amount of data scanned (\$0.002/GB) in processing an S3 Select query and the amount of data returned (\$0.0007/GB). 
The cost for data return depends on the selectivity of the query. 
Data scan and transfer cost is typically a major component in overall query cost when S3 Select is in use.

\spara{Network request cost}
Issuing HTTP requests to Amazon S3 is charged based on the request type and the number of requests. We consider only the cost of HTTP GET requests (\$0.0004 per 1,000 requests) as this is the only request type we use. 
This cost is paid for both S3 Select requests and conventional table read requests. 

\spara{Computation cost}
We used EC2 memory-optimized instances for our experiments. The query execution time is measured in seconds and used to calculate the computation cost based on the hourly price of the host EC2 instance (e.g., r4.8xlarge instances costs \$2.128 per hour). 
The computation cost is another significant component of the overall query cost.

\section{Database Testbed: PushdownDB} 
\label{sec:testbed}

In order to explore how S3 Select can be leveraged to improve query performance and/or reduce cost, we implemented a bare-bone row-based DBMS testbed, called \name. 
We concluded that modifying a commercial multi-node DBMS (e.g., Presto) would be a prohibitive amount of work.  Instead, we implemented \name which has a minimal optimizer and an executor that enables the experiments in this paper.  

We made a reasonable effort to optimize \name's performance.  
While we could not match the performance of the more mature Presto system on all queries, we obtained competitive performance, as shown in Section~\ref{sec:tpch}. 
The source code of \name is available on github at \url{
https://github.com/yxymit/s3filter.git}, and is implemented in a mixture of C++ and Python.

\name represent the query plan as a directed acyclic graph and executes queries in either serial or parallel mode. In the serial mode, a single CPU executes one operator at a time. 
In the parallel mode, each operator executes in parallel using multiple Python processes and passes batches of tuples from producer to consumer using a queue. 
Most operators achieve better performance in the parallel mode, but some operators can benefit from serial mode. A projection followed by a filter, for example, can demonstrate better performance when run in the same process, because this avoids inter-process data transfers.
Most queries in this paper are executed in a mixture of the two modes. 

A few performance optimizations have been built into \name. 
For example, \name does not use SSL as we expect analytics workloads would typically be run in a secure environment. 
Also, \name uses the Pandas library~\cite{mckinney2011pandas} to represent collections of tuples as data frames, generating a significant performance advantage over implementing tuple processing in plain Python.

\spara{Experimental Setup} \label{ssec:config} 
Experiments in this paper are performed on an r4.8xlarge EC2 instance, which contains 32 physical cores, 244~GB of main memory, and a 10~GigE network. The machine runs Ubuntu 16.04.5 LTS. 
\name is executed using Python version 2.7.12.

Unless otherwise stated, all experiments use the same 10~GB TPC-H dataset in CSV format. We will also report Parquet experiments in Section~\ref{sec:parquet}. To facilitate parallel processing, each table is partitioned into multiple objects in S3. 
The techniques discussed in this paper do not make any assumptions about how the data is partitioned. 
During execution, \name spawns multiple processes to load data partitions in parallel.

\section{Filter} \label{sec:filter}

This section discusses how \name accelerates filter operators using S3 Select. 
Given that it is straightforward to pushdown a where clause to S3, we focus on the more interesting problem of supporting indexing using S3 Select. 

\subsection{Indexing with S3 Select}
\label{ssec:indexing}

Both hash indexes and tree-based indexes are widely used in database systems. 
Neither implementation, however, is a good fit for a cloud storage environment because a single index lookup requires multiple accesses to the index. 
This causes multiple S3 requests that incur long network delays. 
To avoid this issue, we designed an index table that is amenable to the filtering available in S3 Select.

An index table contains the values of the columns to be indexed, as well as the byte offsets of indexed records in that table. Specifically, an index table has the following schema (assuming the index is built on a single column).

\small
\begin{verbatim}
|value|first_byte_offset|last_byte_offset|
\end{verbatim}

\normalsize

Accessing a table through an index comprises two phases. 
In phase 1, the predicate on the indexed columns is sent to the index table using an S3 Select request. Then the byte offsets of selected rows are returned to the compute server. In phase 2, the byte offsets are used to directly load the corresponding rows from the data table, by sending an HTTP request for each individual row. Note that accesses in the second phase do not use S3 Select and therefore do not incur the associated extra cost.

\subsection{Performance Evaluation}






\begin{figure}[t!]
    \centering
    \subfloat[Runtime]{
        \includegraphics[width=0.98\columnwidth]{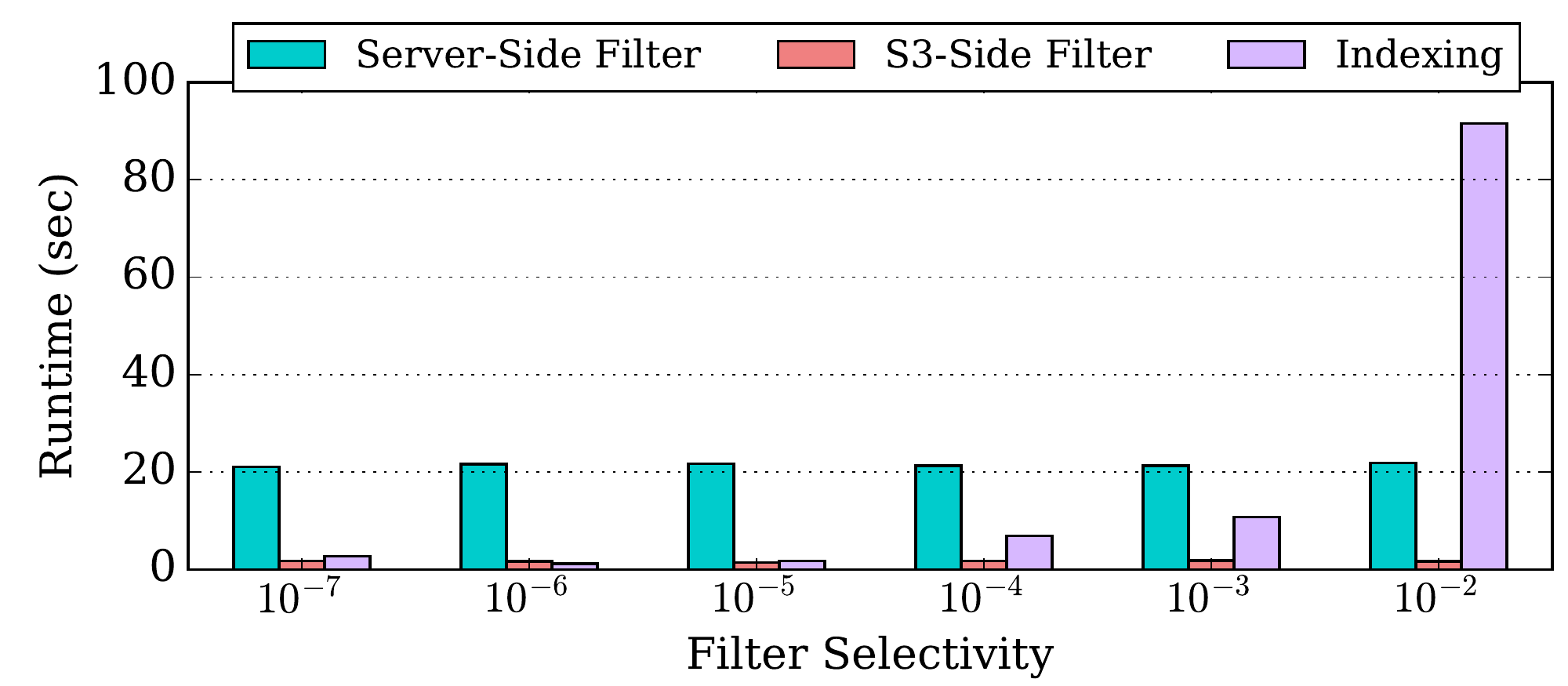}
        \label{fig:filter-rt}
    }
    \\  \vspace{-.05in}
    \subfloat[Cost]{
     \includegraphics[width=0.98\columnwidth]{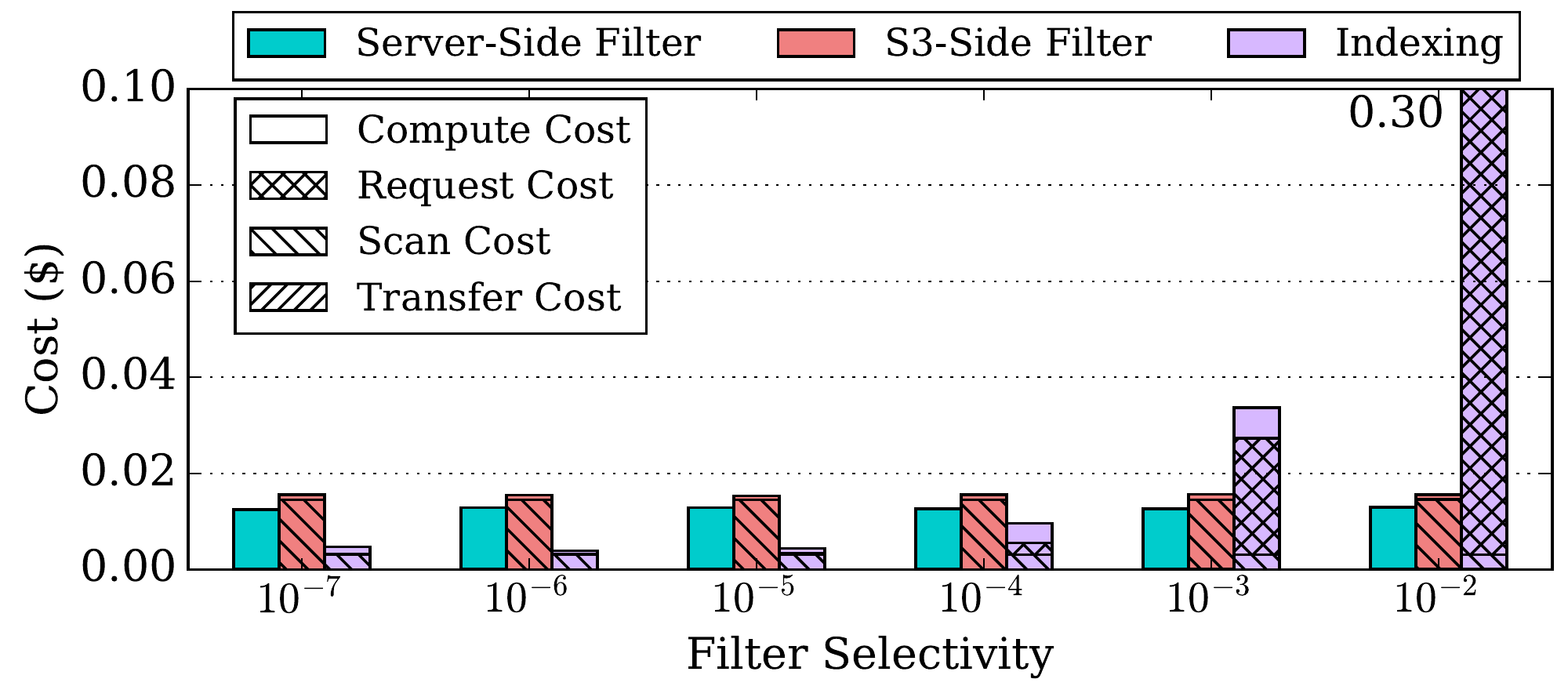}
        \label{fig:filter-cost}
    }
    \caption{
\textbf{Filter algorithms} --- \normalfont{Performance and cost of three filtering strategies as the filter selectivity changes.}
\vspace{-.1in}
    }
    \label{fig:filter}
\end{figure}

Figure~\ref{fig:filter} shows the runtime and cost of different filtering algorithms as the filter selectivity increases from $10^{-7}$ to $10^{-2}$.
\textit{Server-side filter} loads the entire table from S3 and performs filtering on the compute node. \textit{S3-side filter} sends the filtering predicate to S3 in an S3 Select request. \textit{S3-side indexing} uses the index table implementation. 

The performance improvement (Figure~\ref{fig:filter-rt}) from server-side filter to S3-side filter is a dramatic $10\times$ and remains stable as selectivity changes in the specified range.
S3-side indexing has similar performance as S3-side filter when the filter is highly selective, but the performance of indexing degrades as the the filter selects more than $10^{-4}$ of the rows.
In this case, more rows are returned and most of the execution time is spent requesting and receiving individual byte ranges from the data table. 
Although these requests are sent in parallel, they incur excessive CPU computation that become a performance bottleneck.

The cost (Figure~\ref{fig:filter-cost}) of each run is broken down into four components: compute cost, S3 request cost, S3 data scan cost, and data transfer cost.
Each component is denoted using a different type of hash marks. 
Overall, S3-side filter is 24\% more expensive than server-side filter.
Most of the cost of S3-side filter is due to S3 data scanning and loading, while most of the cost of server-side filter is due to computation.
S3-side indexing is cheaper than server-side filter by $2.7\times$ when the filter is very selective, because the index table redues the amount of data being scanned and transferred.
As the filter passes more data, however, the cost of indexing grows rapidly due to increasing HTTP requests.

In conclusion, S3-side indexing is the best approach with highly selective queries, whereas S3-side filter achieves a good balance between performance and cost for queries with any selectivity.


\section{Join} \label{sec:join}

S3 Select does not support pushing a join operator in its entirety into S3. 
This section shows how \name breaks down a join to partially leverage S3 Select. 

It is inherently difficult to take advantage of pushdown processing for joins. 
The two tables to be joined are typically partitioned across multiple S3 objects so that data can be loaded in parallel. 
If the two tables are not partitioned on the join key, implementing a join operator requires shuffling data among different partitions, which is challenging to support at the storage layer. 
\name supports joining tables not partitioned on the join key, as we describe next.

We limit our discussion to hash joins implemented using two phases: the \textit{build phase} loads the smaller table in parallel and sends each tuple to the appropriate partition to build a hash table; the \textit{probe phase} loads the bigger table in parallel and sends the tuples to the correct partition to join matching tuples by probing the hash table.

\subsection{Join Algorithms}

\name supports three join algorithms: \textit{Baseline Join}, \textit{Filtered Join}, and \textit{Bloom Join}. 
These algorithms leverage S3 Select in different ways.

In baseline join, the server loads both tables from S3 and executes the hash join locally, without using S3 Select. Filtered join pushes down selection and projection using S3 Select, and executes the rest of the query in the same way as baseline join. We will not discuss these two algorithms in detail due to their limited use of S3 Select.

In this section, we focus our discussion on Bloom join: after the build phase, a Bloom filter is constructed based on the join keys in the first table; the Bloom filter is then sent as an S3 Select request to load a filtered version of the second table.  In other words, rows that do not pass the Bloom filter are not returned.

\subsubsection{Bloom Filter}

A Bloom filter~\cite{bloom1970space} is a probabilistic data structure that determines whether an element exists in a set or not. 
A Bloom filter has no false negatives but may have false positives. If a Bloom filter returns false, the element is definitely not in the set; if a Bloom filter returns true, the element may be in the set. Compared to other data structures achieving the same functionality, a Bloom filter has the advantage of high space efficiency.

A Bloom filter contains a \textit{bit array} of length $m$ (initially containing all 0's) and uses $k$ different \textit{hash functions}. 
To add an element to a Bloom filter, the $k$ hash functions are applied to the element. 
The output of each hash function is a position in the bit array, which is then set to 1. 
Therefore, at most $k$ bits will be set for each added element. 
To query an element, the same $k$ hash functions are applied to the element. 
If the corresponding bits are all set, then the element may be in the set; otherwise, the element is definitely not in the set. 
The false positive rate of a filter is determined by the size of the set, the length of the bit array, and the hash functions are being used.

Universal hashing~\cite{universal_hashing} is a good candidate for our Bloom filter approach as it requires only arithmetic operators (which S3 Select supports) and represents a family of hash functions.
The hash functions that we use can be generalized as:

\vspace{-.15in}
\begin{align*}
h_{a,b}(x) = ((a \times x + b) \bmod n) \bmod m
\end{align*}

Where $m$ is the length of the bit array and $n$ is a prime $\ge$ $m$. $a$ and $b$ are random integers between $0$ and $n - 1$, where $a \ne 0$.


Given a desired false positive rate $p$, the number of hash functions $k_p$ and the length of the bit array $m_{p}$ are determined by the following formulas \cite{scalable_bloom_filters}, where $s$ is the number of elements in the set.

\vspace{-.1in}
\begin{align*}
k_{p} = \log_2 \frac{1}{p}, \ \ \ \ \ \ 
m_{p} = s \times \frac{\lvert \ln p \rvert}{(\ln 2) ^ 2}
\end{align*}



\subsubsection{Bloom Join in PushdownDB}


Bloom filters are usually processed using bitwise operators. However, since S3 Select does not support bitwise operators or binary data, an alternative is required that not only represents the bit array but can be tested for the presence of a set bit. In \name, we use strings of 1's and 0's to represent the bit array. The following example shows what an S3 Select query containing a Bloom filter would look like. The arithmetic expression on \texttt{attr} within the \texttt{SUBSTRING} function is the hash function on \texttt{attr}.

\lstset{basicstyle=\ttfamily\scriptsize}
\begin{lstlisting}[caption=Example Bloom filter query,label=lst:bloom-example,captionpos=b]
SELECT
    ...
FROM
    S3Object
WHERE
    SUBSTRING('1000011...111101101',
      ((69 * CAST(attr as INT) + 92) % 97) % 68 + 1, 1
    ) = '1'
\end{lstlisting}


With Bloom join, the first table (typically the smaller one) is loaded with filtering and projection pushed to S3. 
The returned tuples are used to construct both the Bloom filter and the hash tables. 
The Bloom filter is then sent to S3 to load the second table.
The returned tuples then probe the hash table to finish the join operation. 

The current implementation of Bloom join supports only integer join attributes. 
This is because the hash functions only support integer data types at present. 
A more general support for hashing in the S3 Select API would enable Bloom joins on arbitrary attributes. 
In fact, while algorithms exist for hashing variable-length strings, they require looping and/or array processing operators that are not currently available to S3 Select queries. 
Additionally, since the bit array is represented using 1 and 0 characters, the bit array is much larger than it would be if S3 Select had support for binary data or bitwise operators to test the presence of a set bit. 
We believe that extending the S3 Select interface in this fashion would be beneficial in our Bloom join algorithm, and perhaps elsewhere.

\subsection{Performance Evaluation}

We compare the runtime and cost of the three join algorithms: baseline, filtered, and Bloom joins. 
Our experiments use the customer and orders tables from the TPC-H benchmark with a scale factor of 10. The following SQL query will be used for evaluation.
Our experiments will sweep two parameters in the query, \texttt{upper\_c\_acctbal} and \texttt{upper\_o\_orderdate}, with their default values being $-950$ and None, respectively.

\lstset{basicstyle=\ttfamily\scriptsize}
\begin{lstlisting}[caption=Synthetic join query for evaluation,label=lst:join-query,captionpos=b]
        SELECT
            SUM(O_TOTALPRICE)
        FROM
            CUSTOMER, ORDER
        WHERE
            O_CUSTKEY = C_CUSTKEY AND
            C_ACCTBAL <= upper_c_acctbal AND
            O_ORDERDATE < upper_o_orderdate
\end{lstlisting}

\subsubsection{Customer Selectivity}

\begin{figure}[t]
	\centering
	\subfloat[Runtime]{
		\includegraphics[width=0.98\columnwidth]{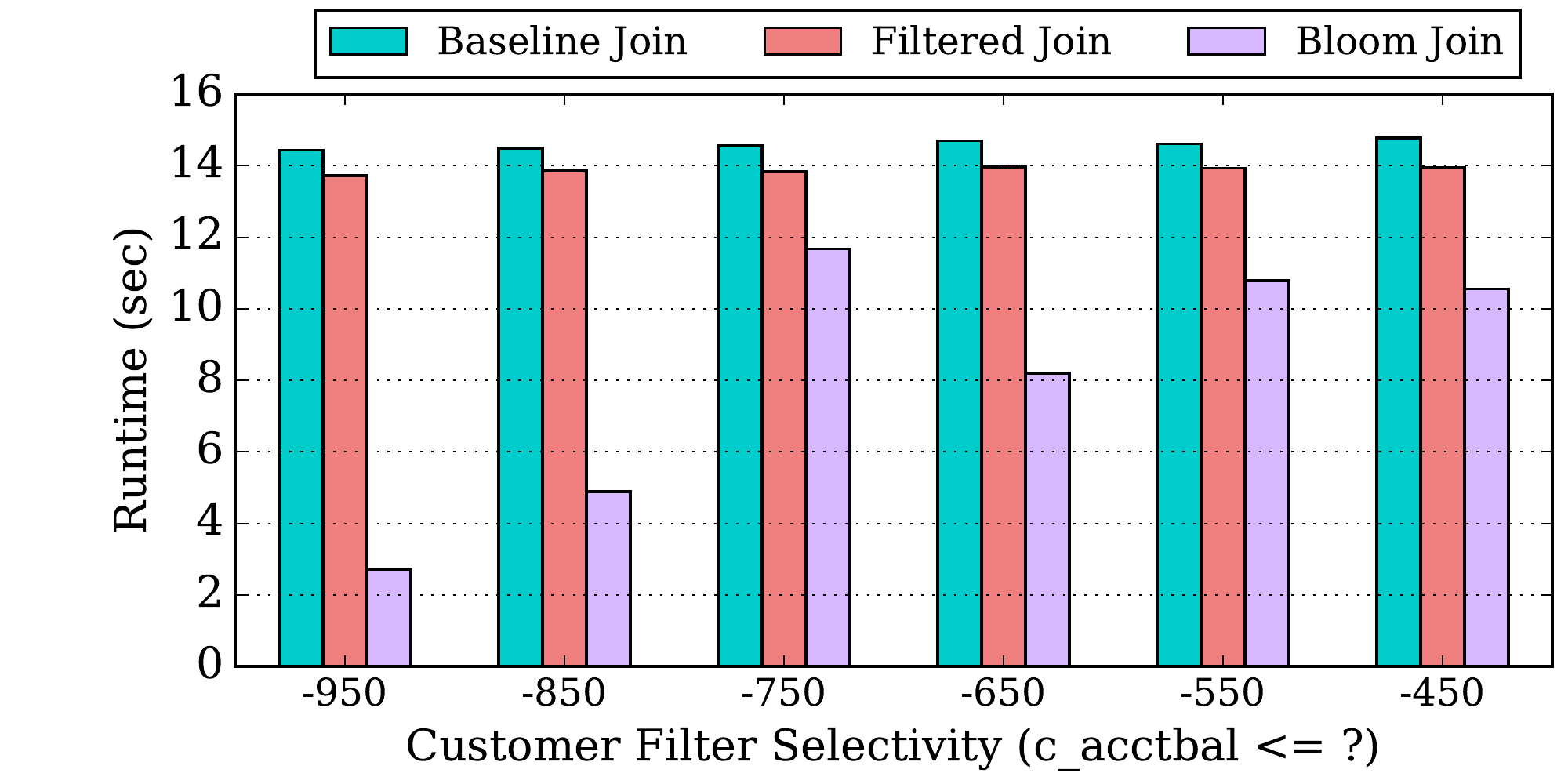}
		\label{fig:join-aval-rt}
	} \\ \vspace{-.05in}
	\subfloat[Cost]{
		\includegraphics[width=0.98\columnwidth]{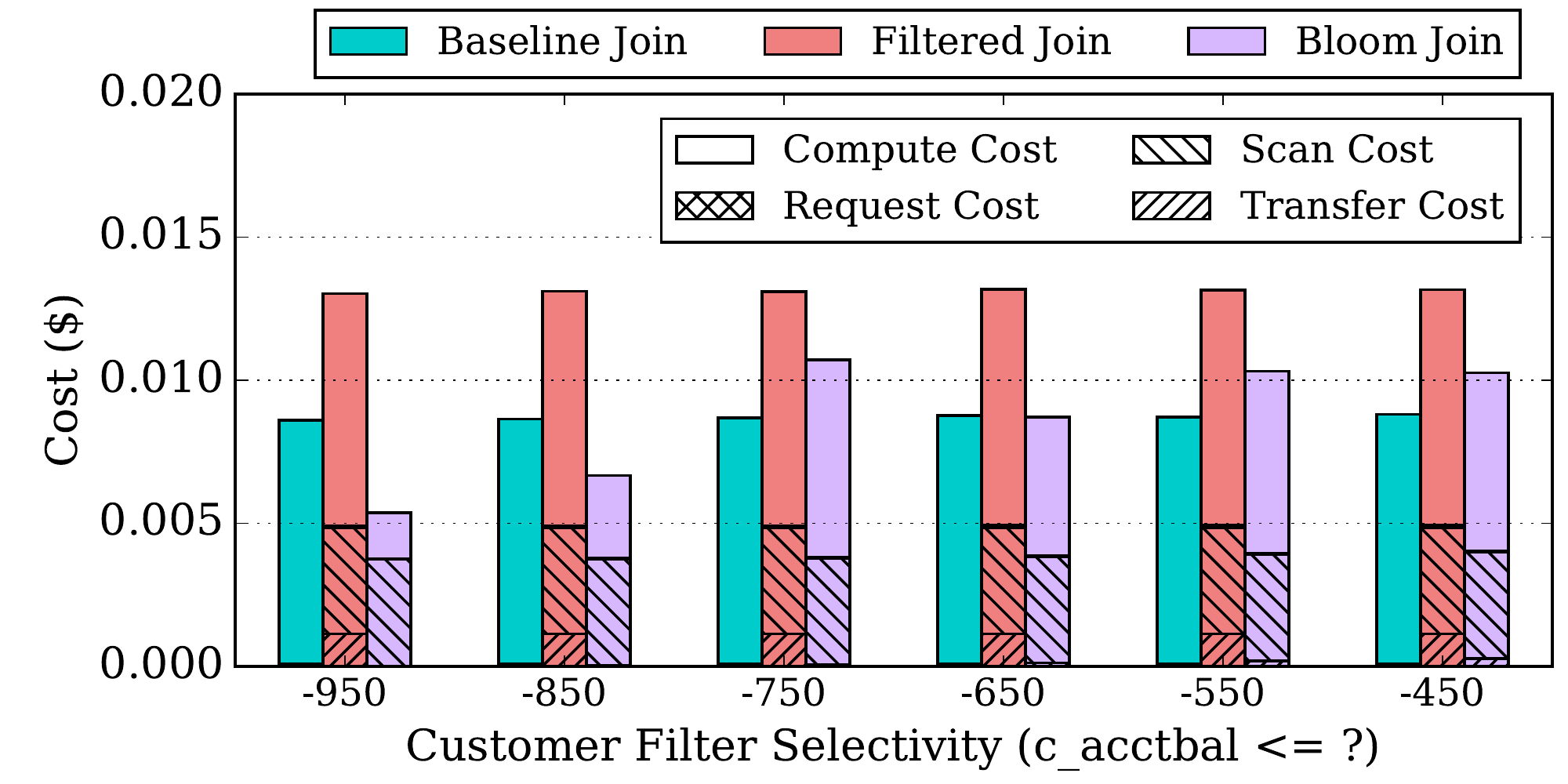}
		\label{fig:join-aval-cost}
	} \\ 
	\caption{
		\textbf{Customer selectivity} --- \normalfont{Performance and cost when varying customer table selectivity.
		}
	}
	\vspace{-.1in}
	\label{fig:join-aval}
\end{figure}

This experiment sweeps selectivity on the customer table by varying the value of \texttt{upper\_c\_accbal} from -950 to -450, meaning relatively small numbers of tuples are returned from the customer table. 
For this experiment, the orders table selectivity is fixed at `None' (all rows are returned). The false positive rate for the Bloom filter is $0.01$.

Figure~\ref{fig:join-aval} shows the runtime and cost of different join algorithms as the selectivity on the customer table changes. 
Baseline and filtered joins perform similarly, which is expected since they only apply selection to the smaller customer table and load the entire orders table, which incurs the same large amount of network traffic. 
Bloom join performs significantly better than either as the high selectivity on the first table is encapsulated by the Bloom filter, which significantly reduces the number of returned rows for the larger orders table. 
As the predicate on the customer table becomes less selective, Bloom join’s performance degrades as fewer records are filtered by the Bloom filter. 
Bloom join is cheaper than the other two algorithms with high selectivity, although the cost advantage is not as significant as the runtime advantage.

It is important to note that the limit on the size of S3 Select's SQL expressions is 256KB. In this example, if the selectivity on the customer table is low, the required Bloom filter needs to be bigger and thus may exceed the size limit. 
\name detects this case and increases the false positive rate for the Bloom filter to ensure this limit is not exceeded. In the case where the best achievable false positive rate cannot be less than 1, \name falls back to not using a Bloom filter at all, resulting in an algorithm similar to a filtered join. 
However, there is one difference between the degraded Bloom join and a filtered join: in the Bloom join, the two table scans happen serially, since the decision to revert to filtered join is made only after the first table is loaded. 
The original filtered join algorithm can load the two tables in parallel, thereby performing better than a degraded Bloom join.

\subsubsection{Orders Selectivity}

\begin{figure}[t!]
	\centering
	\subfloat[Runtime]{
		\includegraphics[width=0.98\columnwidth]{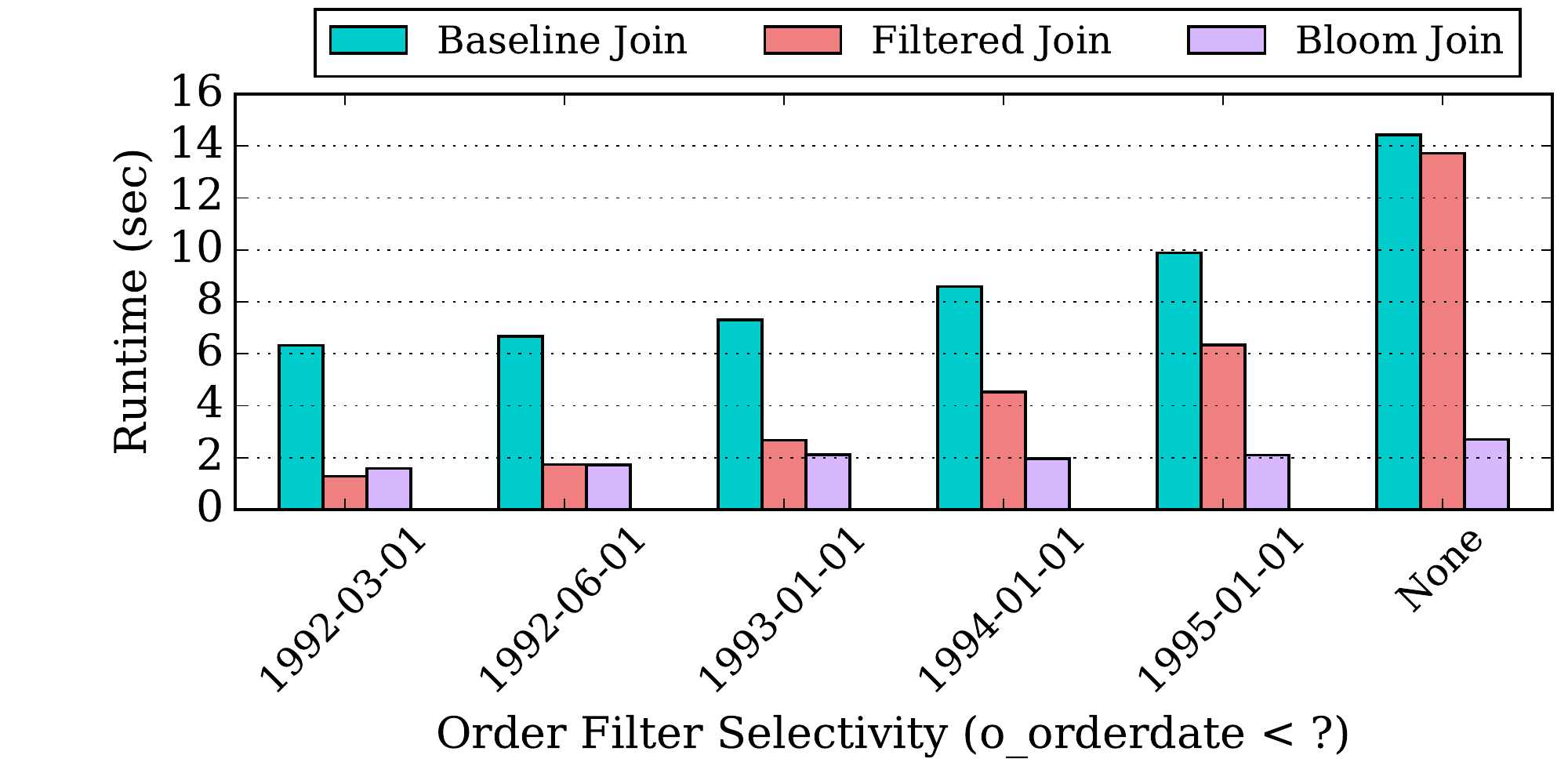}
		\label{fig:join-bval-rt}
	} \\ \vspace{-.05in}
	\subfloat[Cost]{
		\includegraphics[width=0.98\columnwidth]{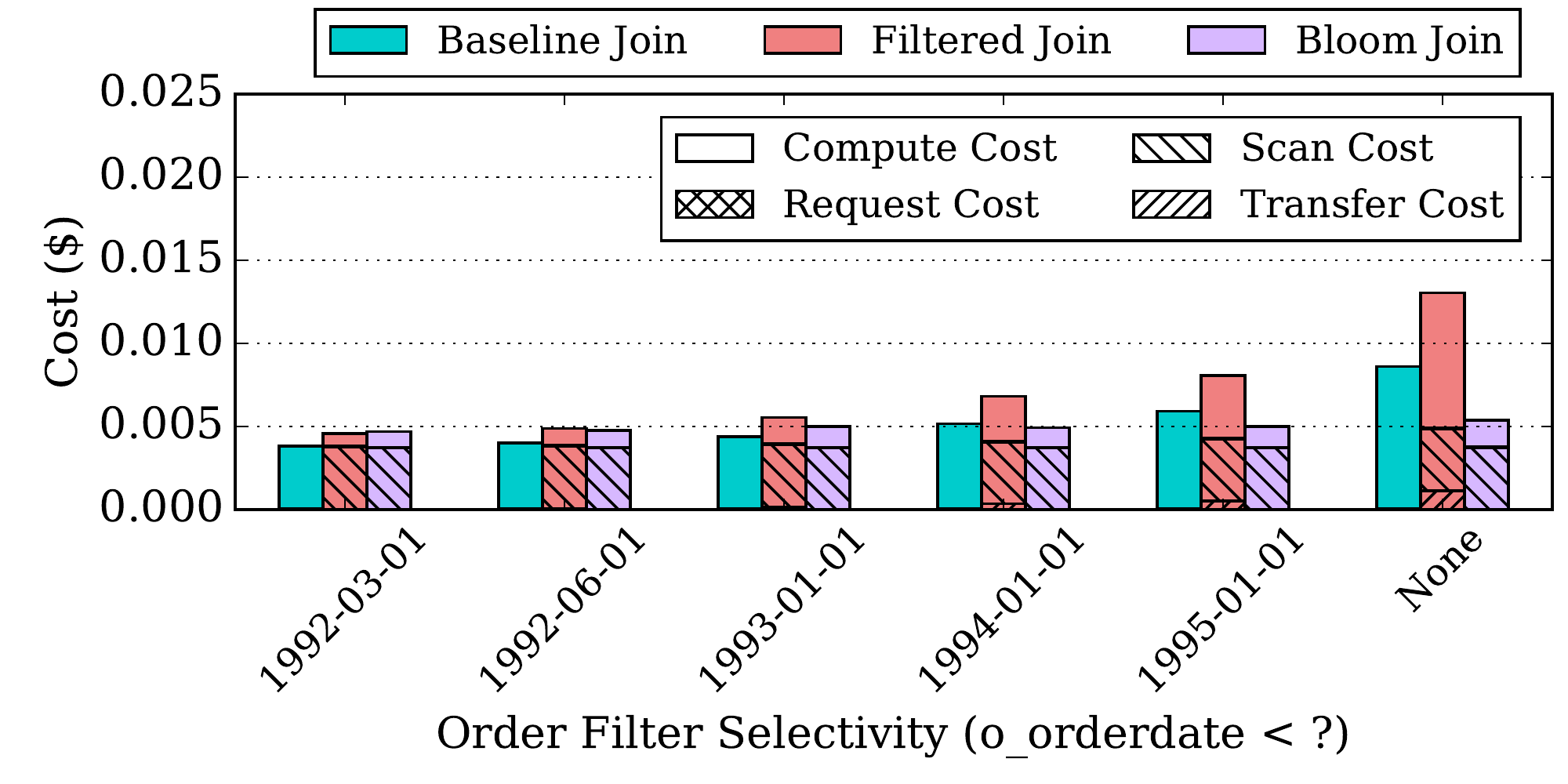}
		\label{fig:join-bval-cost}
	} \\ 
	\caption{
		\textbf{Orders selectivity} --- \normalfont{Performance and cost when varying the orders table selectivity.
		}
	} 
	\vspace{-.1in}
	\label{fig:join-bval}
\end{figure}

This experiment fixes customer table selectivity at -950 (highly selective) and the false positive rate for the Bloom filter at 0.01. 
The selectivity for the orders table is swept from high to low by limiting records returned from the orders table by sweeping \texttt{upper\_o\_orderdate} in the range of [`1992-03-01', `1992-06-01', `1993-01-01', `1994-01-01', `1995-01-01', None].

The results are shown in Figure~\ref{fig:join-bval}. 
Filtered join performs significantly better than baseline join when the filter on the orders table is selective. 
The performance advantage disappears when the filter becomes less selective. 
Bloom join performs better and remains fairly constant as the number of records returned from the orders table remains small due to the Bloom filter. 
The cost of Bloom join is either comparable or cheaper than the alternatives.

\subsubsection{Bloom Filter False Positive Rate}

This experiment fixes both customer table selectivity and orders table selectivity at -950 and `None', respectively. The false positive rate for Bloom Join is swept from low to high to low using the rates [0.0001, 0.001, 0.01, 0.1, 0.3, 0.5].

\begin{figure}[t!]
	\centering
	\subfloat[Runtime]{
		\includegraphics[width=0.98\columnwidth]{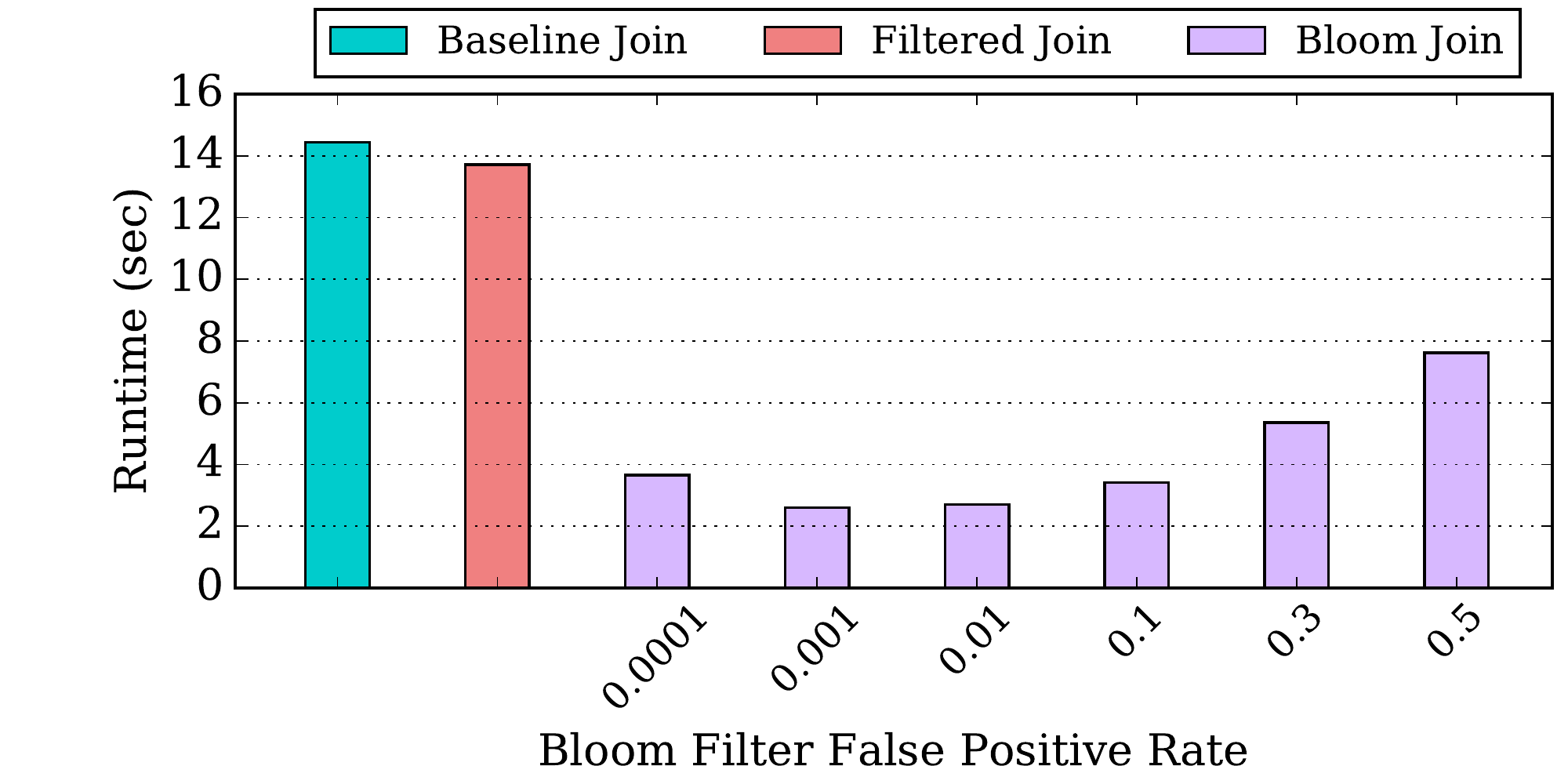}
		\label{fig:join-fp_rate-rt}
	} \\ \vspace{-.05in}
	\subfloat[Cost]{
		\includegraphics[width=0.98\columnwidth]{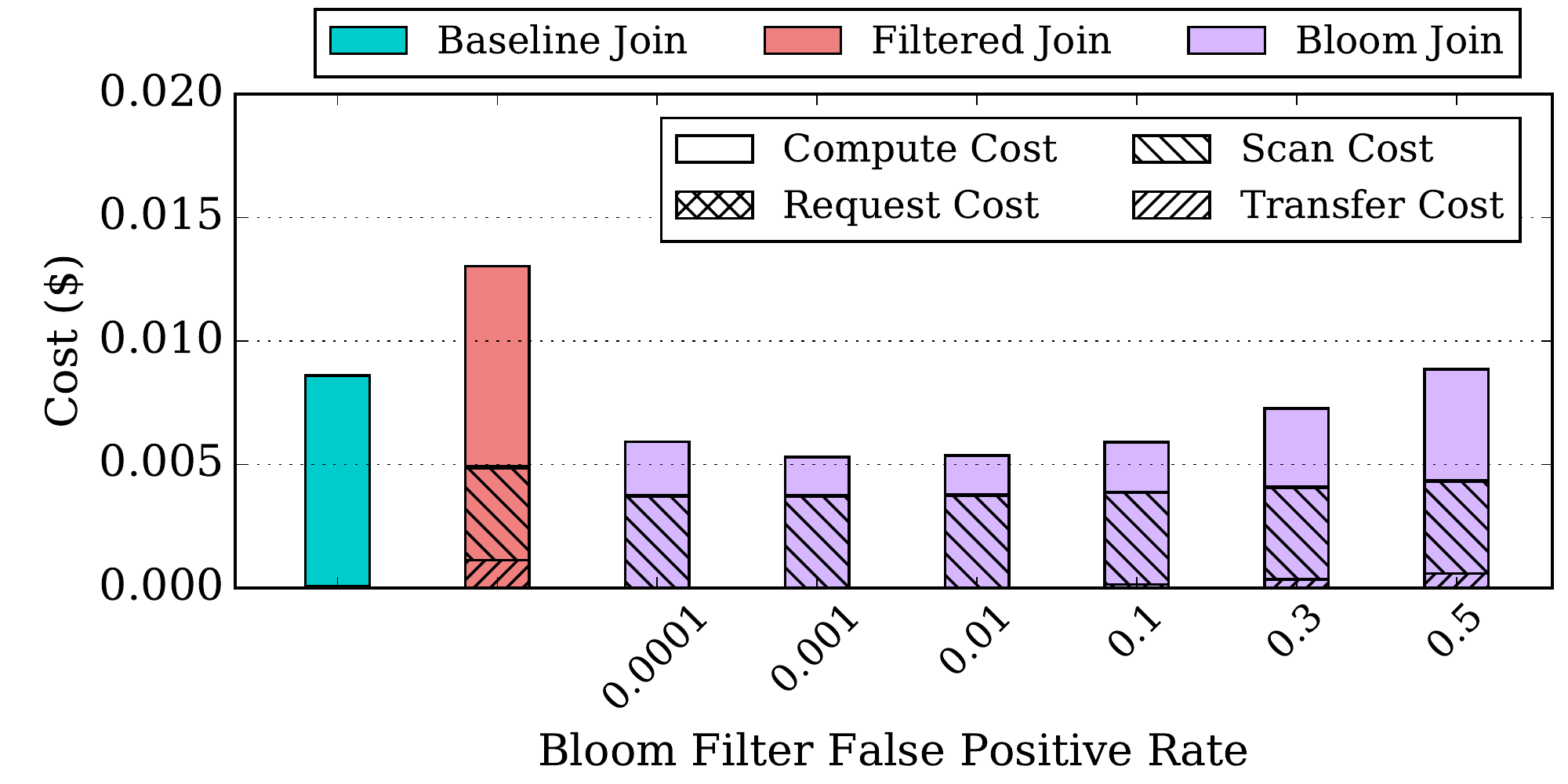}
		\label{fig:join-fp_rate-cost}
	} \\ 
	\caption{
		\textbf{Bloom filter false positive rate} --- \normalfont{Performance and cost when varying the Bloom filter false positive rate. 
		}
	} 
	\vspace{-.15in}
	\label{fig:join-fp_rate}
\end{figure}

Figure~\ref{fig:join-fp_rate} shows the runtime and cost of baseline and filtered join as well as Bloom join with different false positive rates. 
We can see that the best performance and cost numbers can be achieved when the false positive rate is 0.01. 
When the false positive rate is low, the Bloom filter is large in size, increasing the computation requirement in S3 Select. 
When the false positive rate is high, the Bloom filter is less selective, meaning more data will be returned from S3. 
A rate of 0.01 strikes a balance between these two factors.

\section{Group-By} \label{sec:groupby}

The current S3 Select supports simple aggregation on individual attributes but not with a group-by clause. 
Pushing a group-by aggregation to S3 is desirable as it can significantly reduce network traffic. 
In this section, we explore designs of group-by algorithms that leverage S3 Select. 







Group-by can be performed at the server-side by loading all data from S3 directly (\textit{Server-side group-by}) or loading S3 data using a predicate (\textit{Filtered group-by}). Both implementations are straightforward. Therefore, we focus our discussion on two other algorithms that are less obvious to implement but deliver better performance --- 
\textit{S3-side group-by} and \textit{Hybrid group-by}.

\subsection{S3-Side Group-By}
\label{sec:s3-groupby}

The S3-side group-by algorithm pushes the group-by logic entirely into S3 and thus minimizes the amount of network traffic.
We use the following query to demonstrate how the algorithm works. 
It computes the total account balance for each nation in the customer table.

\lstset{basicstyle=\ttfamily\scriptsize}
\begin{lstlisting}[caption=Example group-by query,label=lst:groupby,captionpos=b]
        SELECT c_nationkey, sum(c_acctbal)
        FROM customer
        GROUP BY c_nationkey;
\end{lstlisting}

The first phase of execution collects the values for the groups in the group-by clause. 
For the example query, we need to find the unique values of \texttt{c\_nationkey}. 
This is accomplished by running a projection using S3 Select to return only the \texttt{c\_nationkey} column (i.e., \texttt{SELECT c\_nationkey FROM customer}). The compute node then finids unique values in the  column.

In the second phase of execution, \name requests S3 to perform aggregation for each individual group that the first phase identified. 
For example, if the unique values of \texttt{c\_nationkey} are 0 and 1, then the following query will be sent to S3 in phase 2.

\lstset{basicstyle=\ttfamily\scriptsize}
\begin{lstlisting}[caption=Phase 2 of S3-side group-by,label=lst:s3-groupby-p2,captionpos=b]
        SELECT sum(CASE WHEN c_nationkey = 0
                   THEN c_acctbal ELSE 0 END),
               sum(CASE WHEN c_nationkey = 1
                   THEN c_acctbal ELSE 0 END)
               ...
        FROM customer;
\end{lstlisting}

The first and second returned numbers are the total customer balance for \texttt{c\_nationkey} = 0 and 1, respectively. 
The number of columns in the S3 Select response equals the number of unique groups multiplied by the number of aggregations. 
The query execution node converts the results into the right format and returns them to the user.

\subsection{Hybrid Group-By}
\label{sec:hybrid-groupby}

In practice, many data sets are highly skewed, with a few large groups containing the majority of rows, and many groups containing only a few rows. 
For these workloads, \textit{S3-side group-by} will likely deliver bad performance since the large number of groups leads to long S3 Select queries. 
To solve this problem, we propose a hybrid group-by algorithm that distinguishes groups based on their size. 
\textit{Hybrid group-by} pushes the aggregation on large groups to S3, thus eliminating the need for transferring large amounts of data. Small groups, on the other hand, are aggregated by the query execution nodes.

Similar to S3-side group-by, hybrid group-by also contains two phases. 
In the first phase, however, hybrid group-by does not scan the entire table, but only a sample of rows as they are sufficient to capture the populous groups. 
In particular, \name scans the first 1\% of data from the table. 

\lstset{basicstyle=\ttfamily\scriptsize}
\begin{lstlisting}[caption=Phase 2 of hybrid group-by,label=lst:hybrid-groupby-p2,captionpos=b]
    Q1: SELECT sum(CASE WHEN c_nationkey = 0
                   THEN c_acctbal ELSE 0 END)             
        FROM customer;

    Q2: SELECT c_nationkey, c_acctbal
        FROM customer
        WHERE c_nationkey <> 0 

\end{lstlisting}

Listing~\ref{lst:hybrid-groupby-p2} shows the S3 Select query for the second phase of hybrid group-by. 
Two queries are sent to S3. Q1 runs remote aggregation for the large groups (in this example, \texttt{c\_nationkey} = 0), similar to the second phase of S3-side group-by. 
Q2 is sent for loading rows belonging to the rest of the groups from S3. Aggregation for these rows is performed locally at the compute node.

\subsection{Performance Evaluation}

We evaluate the performance of different group-by algorithms using synthetic data sets as they allow us to change different parameters of the workload.
We present results with both uniform and skewed group sizes.

\subsubsection{Uniform Group Size}

This section presents experimental results for a dataset with uniform group sizes. Three group-by implementations are included: server-side group-by, filtered group-by, and S3-side group-by. Hybrid group-by will be discussed in detail in the next section. These experiments are performed on a 10~GB table with 20 columns. The first 10 columns contain group IDs and each column contains different numbers of unique groups (from $2$ to $2^{10}$). The group sizes are uniform, meaning each group contains roughly the same number of rows. The other 10 columns contain floating point numbers and are the fields that will be aggregated.

\begin{figure}[t!]
    \centering
    \subfloat[Runtime]{
        \includegraphics[width=0.98\columnwidth]{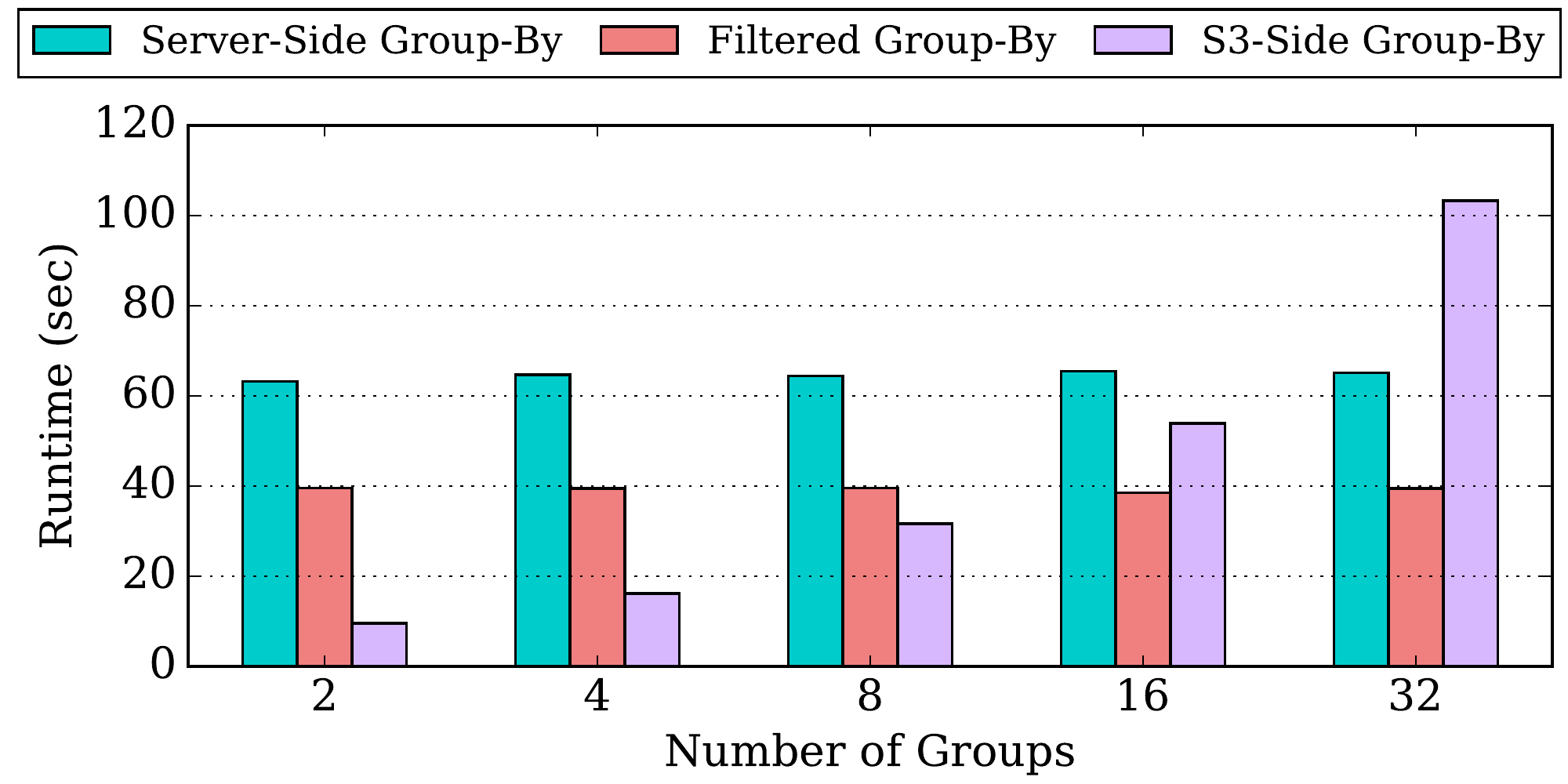}
        \label{fig:ngroups-rt}
    } \\ \vspace{-.05in}
    \subfloat[Cost]{
     \includegraphics[width=0.98\columnwidth]{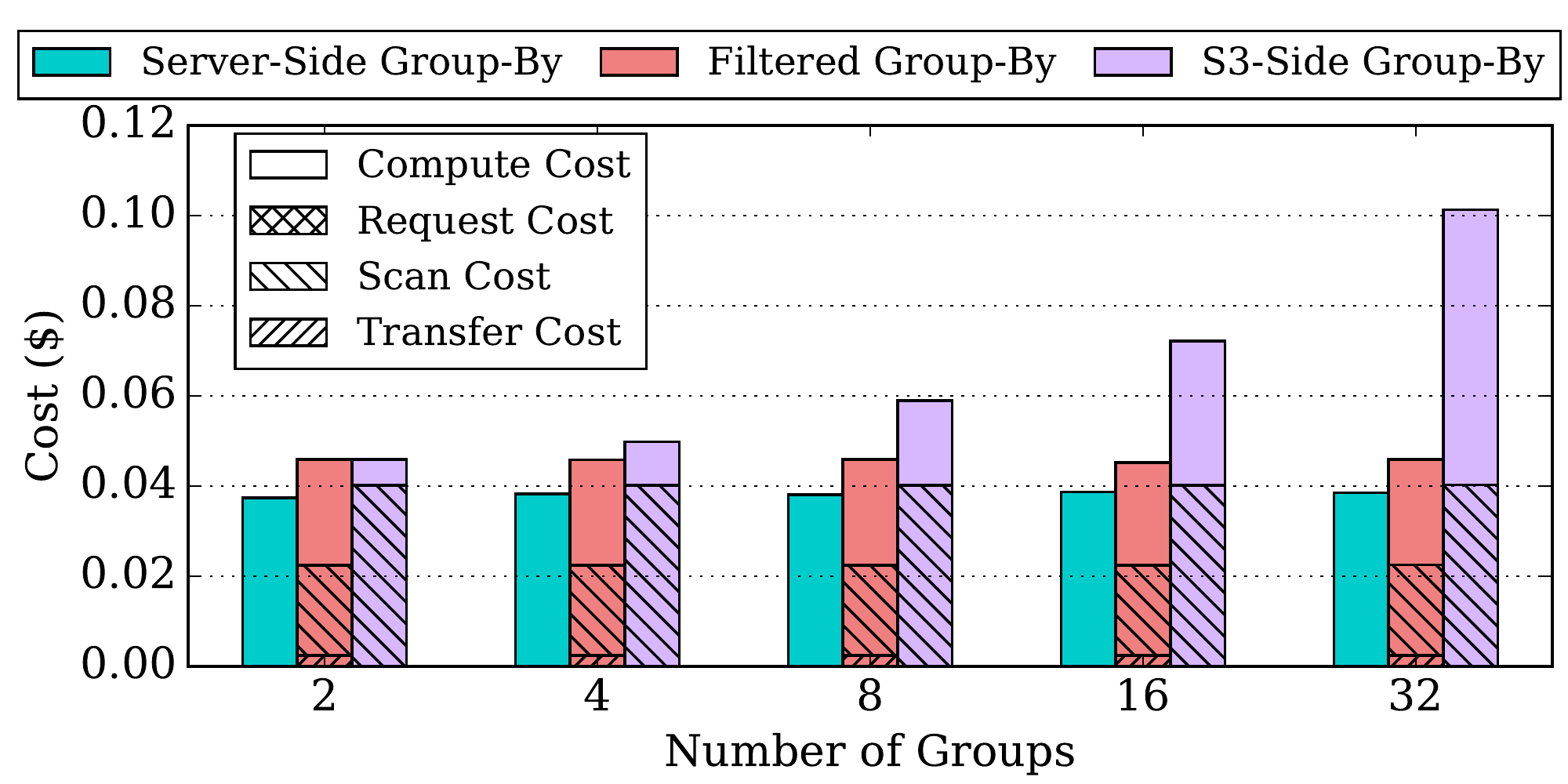}
        \label{fig:ngroups-cost}
    } 
    \caption{
		\textbf{Number of groups} --- \normalfont{Performance and cost as the number of groups increases.}
    } \vspace{-.15in}
    \label{fig:groupby-ngroups}
\end{figure}

Figure~\ref{fig:groupby-ngroups} shows the runtime and cost per query for different group-by algorithms, as the number of groups changes from 2 to 32. Each query performs aggregation over four columns. 
The performance of server-side group-by and filtered group-by does not change with the number of groups, because both algorithms must load all the rows from S3 to the compute node. 
However, filtered group-by loads only the four columns on which aggregation is performed while server-side group-by loads all the columns. This explains the 64\% higher performance of filtered over server-side group-by.
S3-side group-by performs $4.1\times$ better than filtered group-by when there are only a few unique groups. Performance degrades, however, when more groups exist. This is due to the increased computation overhead that is performed by the S3 servers.

Although the three algorithms have relatively high variation in their runtime numbers, the cost numbers are relatively close until eight groups.
The server-side group-by pays more for compute, but the other two algorithms pay more for scanning and transferring S3 data.



\subsubsection{Skewed Group Sizes}

We use a different workload to study the effect of non-uniform group sizes. 
The table contains 10 grouping columns and 10 floating point value columns.
The number of rows within each group is non-uniformly distributed. 
Each grouping column contains 100 groups and the number of rows within each group follows a Zipfian distribution~\cite{gray1994quickly} controlled by a parameter $\theta$. 
A larger $\theta$ means more rows are concentrated in a smaller number of groups. 
For example, $\theta$ = 0 corresponds to a uniform distribution and $\theta = 1.3$ means 59\% of rows belong to the four largest groups.

\begin{figure}[t!]
    \centering
    \includegraphics[width=0.98\columnwidth]{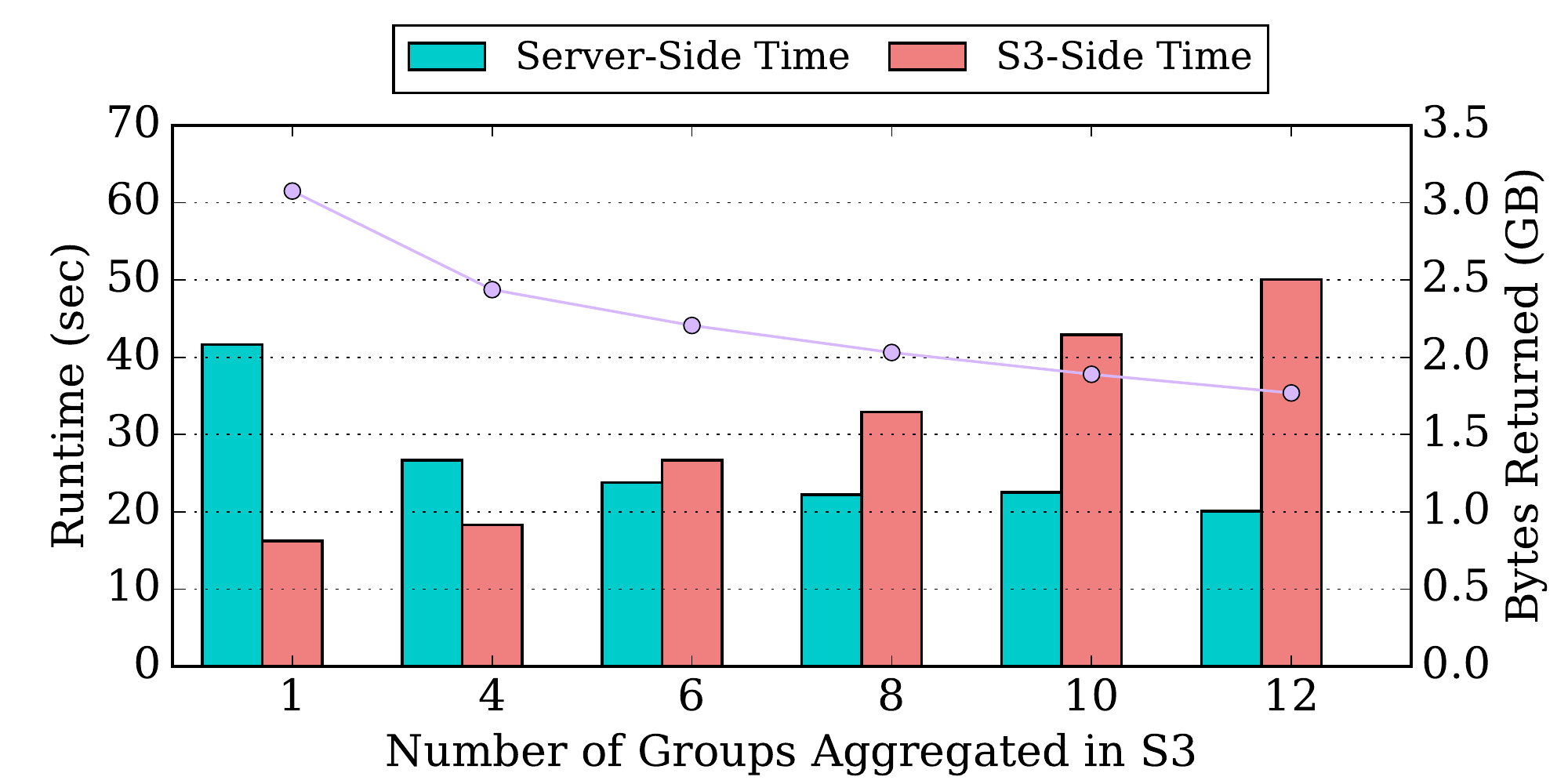} 
    \label{fig:local-vs-remote}
    \caption{
		\textbf{Server- vs. S3-side aggregation in hybrid group-by.}
    } 
    \label{fig:local-vs-remote}
    \vspace{-.2in}
\end{figure}

We first investigate an important parameter in hybrid group-by: how many groups should be aggregated at S3 vs. server side. Figure~\ref{fig:local-vs-remote}  shows the runtime of server-side and S3-side aggregation while increasing the number of groups aggregated in S3. The bars show the runtime and the line shows the number of bytes returned from S3. More S3-side aggregation increases the execution time of the part of query executed at S3, but reduces the amount of data transferred over the network. The final execution time is determined by the maximum of the two bars shown in Figure~\ref{fig:local-vs-remote}.
Overall, having 6 to 8 groups aggregated in S3 offers the best performance.

\begin{figure}[t!]
    \centering
    \subfloat[Runtime]{
        \includegraphics[width=0.98\columnwidth]{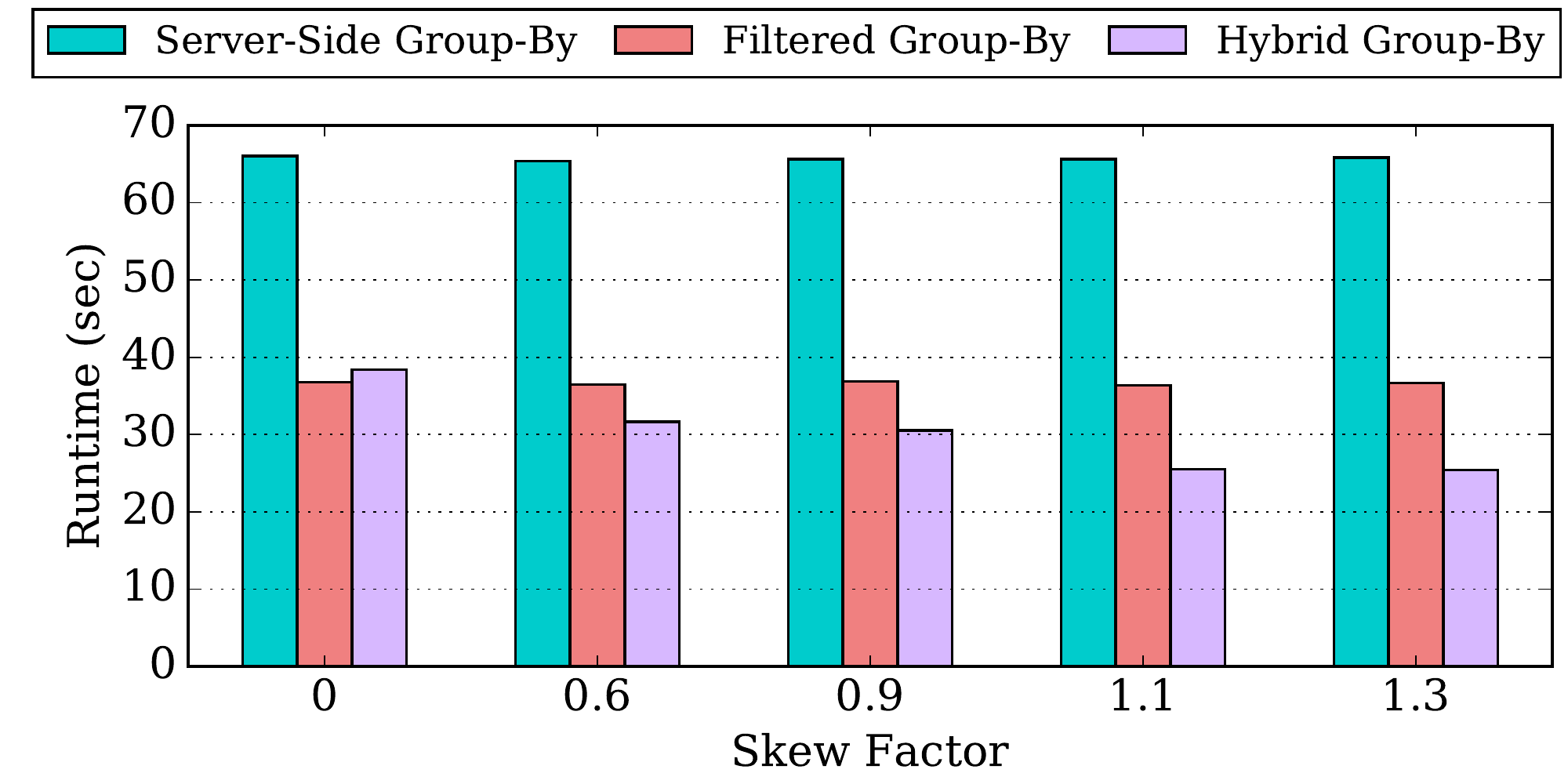}
        \label{fig:skew-rt}
    } \\ \vspace{-.1in}
    \subfloat[Cost]{
     \includegraphics[width=0.98\columnwidth]{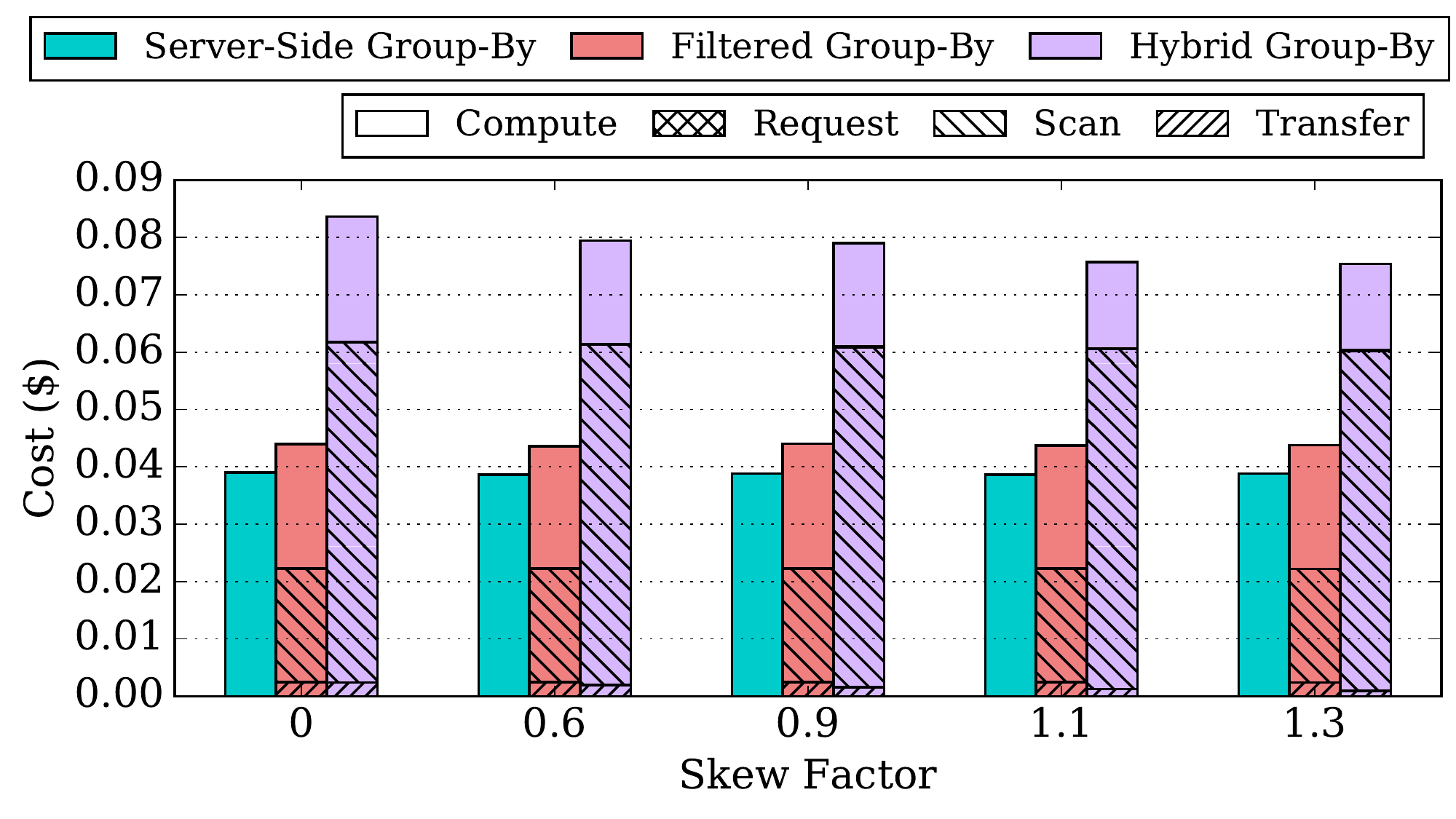}
        \label{fig:skew-cost}
    } \vspace{-.05in}
    \caption{
		\textbf{Data skew} --- \normalfont{Performance and cost with different levels of skew in group sizes.}
    } \vspace{-.2in}
    \label{fig:skew}
\end{figure}

Figure~\ref{fig:skew} shows the performance and cost of three group-by algorithms as the level of skew in group sizes increases. 
Across all levels of skew, the performance and cost of server-side and filtered group-by remain the same. 
In both algorithms, the amount of data loaded from S3 and the computation performed on the server are independent of data distribution. 
When the workload has high skew, the performance advantage of pushing group-by to S3 is evident. 
With $\theta = 1.3$, hybrid group-by performs 31\% better than filtered group-by. 
However, hybrid group-by does not have a cost advantage over the other two algorithms, since it has to scan the table one more time than filtered group-by. 
This extra table scan can be avoided by improving the interface of S3 Select.


\section{Top-k} \label{sec:topk}

Top-K is a common operator that selects the maximum or minimum $K$ records from a table according to a specified expression. 
In this section, we discuss a sampling-based approach that can significantly improve the efficiency of top-K using S3 Select.

\subsection{Sampling-Based Top-K Algorithm} \label{ssec:topk}

The number of records returned by a top-K query, $K$, is typically much smaller than the total number of records in the table, $N$. 
Therefore, transferring the entire table from S3 to the server is inherently inefficient.
We designed a sampling-based two-phase algorithm to resolve this inefficiency: the first phase samples the records from the table and decides what subset of records to load in the second phase; then in the second phase, the query execution node loads this subset of records and performs the top-K computation on it.
We use the following example query for the rest of the discussion. 

\lstset{basicstyle=\ttfamily\scriptsize}
\begin{lstlisting}[caption=Example top-K query,label=lst:topk_query,captionpos=b]
        SELECT *
        FROM lineitem
        ORDER BY l_extendedprice ASC
        LIMIT K;
\end{lstlisting}

During the first phase, we obtain a conservative estimate of a subset that must contain the top-K records. Specifically, 
the system loads a random sample of $S$ ($> K$) records from the S3 and uses the $K^{th}$ smallest \texttt{l\_extendedprice} as the threshold. If the data in the table is random, then the algorithm can simply request the first $S$ records from the table. Otherwise, if the data distribution in the \texttt{l\_extendedprice} column is not random, then a random sample of $S$ records can be obtained by requesting a number of data chunks using random byte offsets from the data table. 
The sampling process guarantees that the top-K records must be below the threshold, since we have already seen $K$ records below the threshold in the sample.
In the second phase, the algorithm uses S3 Select to load records below the threshold.


The number of records returned in the second phase should be between $K$ and $N$. The algorithm then uses a heap to select the top-K records from all returned records.

\subsection{Analysis}
\label{ssec:topk-analysis}

An important parameter in the sampling-based algorithm is the sample size $S$, which is crucial to the efficiency of the algorithm. 
A small $S$ means the second phase will load more data from S3, while a large $S$ means the sampling phase will take significant time. 
The goal of the sampling-based top-K algorithm is to reduce data traffic from S3. 
We can obtain the sample size that minimizes data traffic using the following analysis:

Assume each row contains $B$ bytes, the table contains $N$ rows, and the sample contains $S$ rows. We also assume that only a fraction ($\alpha \leq 1$) of the bytes in a record is needed during the sampling phase, because the expression in the \texttt{ORDER BY} clause does not necessarily require all the columns. We assume the sampling process is uniformly random. The total number of bytes loaded from S3 during the first phase is:

\[ D_1 = \alpha S B \]

The $K^{th}$ record from the sample is selected as the threshold. 
Based on the random sampling assumption, the system loads $KN/S$ records in phase 2. Therefore, the total number of bytes loaded from S3 in phase 2 is:

\[ D_2 = K N B/S \]

The total amount of data loaded from S3 ($D$) is the sum of data loaded during both phases:
    
\vspace{-.15in}
\begin{align*}
D = D_1 + D_2 &= \alpha S B + \frac{K N B}{S}
\end{align*}
\vspace{-.1in}

The value of $S$ that minimizes $D$ can be found by obtaining the derivative of the above expression w.r.t $S$ and equating it to zero. This gives $S = \sqrt{\frac{KN}{\alpha}}$.
Given a fixed table size, a smaller $\alpha$ leads to a bigger $S$. This is because if sampling does not consume significant bandwidth, it is worthwhile to sample more records to improve overall bandwidth efficiency.

\subsection{Performance Evaluation}

In this section, we evaluate the performance of different top-K algorithms using the lineitem table of the TPC-H data set. We use a scale factor of 10, meaning that the lineitem table is $7.25$ GB in size and contains 60 million rows. 
The example query in Listing~\ref{lst:topk_query} is used.

\subsubsection{Sensitivity to Sample Size}
\label{sssec:topk-sample-size}

\begin{figure}[t!]
    \centering
    \subfloat[Runtime]{
        \includegraphics[width=0.98\columnwidth]{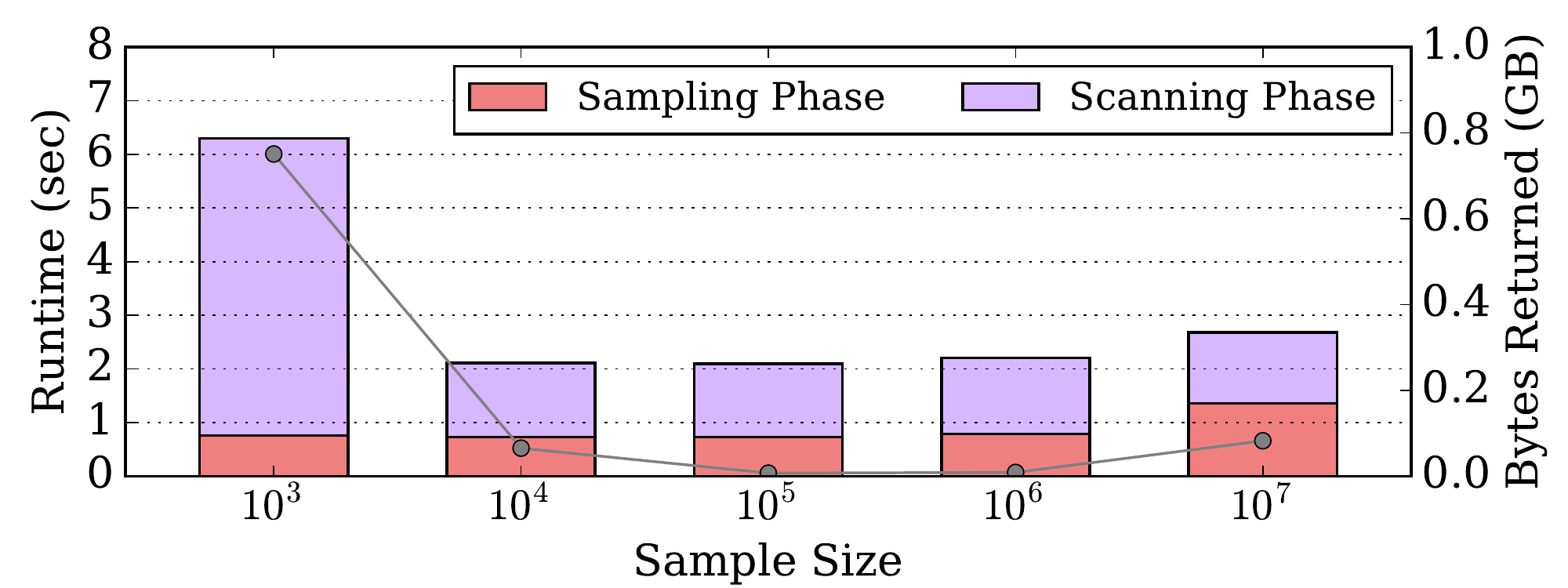}
        \label{fig:topk-sample-rt}
    } \\ \vspace{-.15in}
    \subfloat[Cost]{
     \includegraphics[width=0.98\columnwidth]{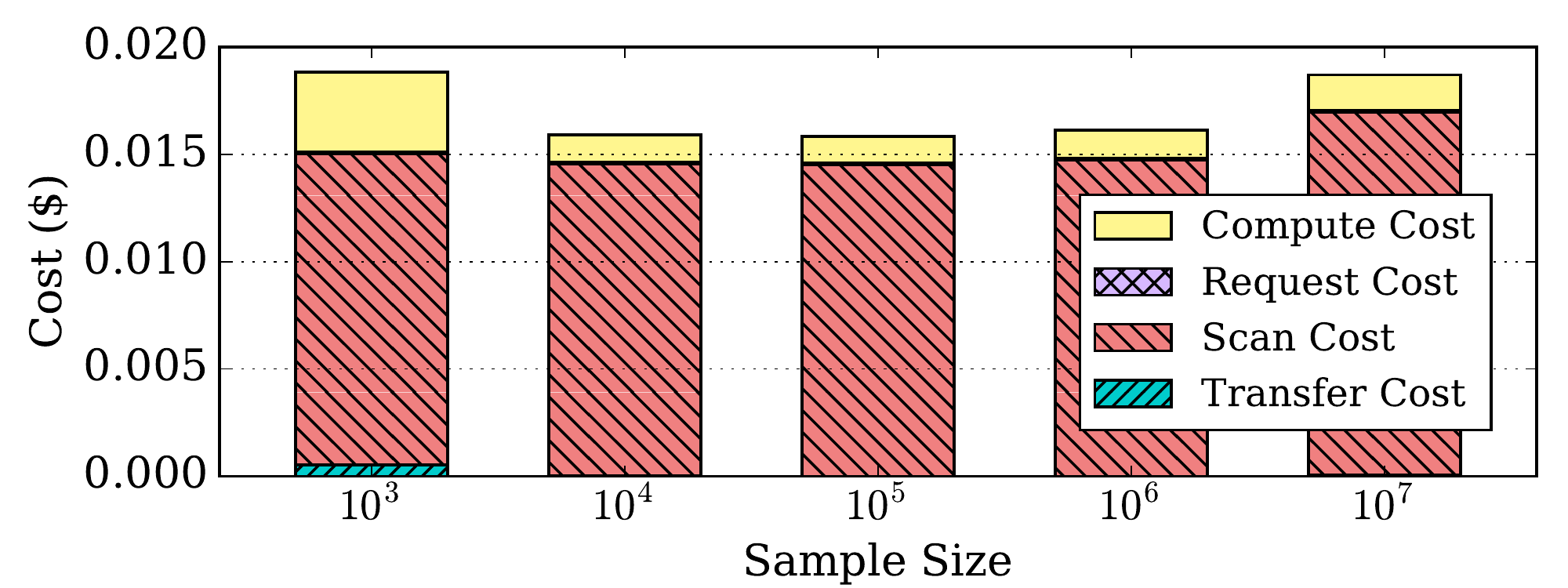}
        \label{fig:topk-sample-cost}
    } \vspace{-.1in}
    \caption{
\textbf{Sensitivity to sample size} --- \normalfont{Performance and cost of the sampling-based top-K as the sample size changes.} } 
    \vspace{-.15in}
    \label{fig:topk-sample}
\end{figure}

We first study how the performance and cost of the sampling-based algorithm change with respect to the sample size $S$. 
For this experiment, we fix $K$ to $100$, and increase $S$ from $10^3$ to $10^7$. Note that $10^3$ is $10$ times $K$, and $10^7$ is $1/6$ of the entire table.

In Figure~\ref{fig:topk-sample-rt}, each bar shows the runtime of a query at a particular sample size. Each bar is broken down into two portions: the sampling phase (phase 1) and the scanning phase (phase 2). The line shows the total amount of data returned from S3 to the server.

As the sample size increases, the execution time of the sampling phase also increases. This is expected because more data needs to be sampled and returned. On the other hand, the execution time of the scanning phase decreases. This is because a larger sample leads to a more stringent threshold, and therefore fewer qualified rows in the scanning phase. The amount of data returned from S3 first decreases due to the dropping S3 traffic in the scanning phase, and later increases due to the growing traffic of the sampling phase. 
Overall, the best performance and network traffic efficiency can be achieved in the middle, when the sample size is around $10^5$.
This result is consistent with our analysis.
According to our model, with $K = 100$, $N = 6\times10^7$, and $\alpha = 0.1$, the calculated optimal sample size $S = \sqrt{\frac{K N}{\alpha}} = 2.4\times10^5$.
The performance of the algorithm is stable in a relatively wide range of values around this optimal $S$.

Figure~\ref{fig:topk-sample-cost} shows the query cost with varying sample size. Most of the cost is due to data scanning, with most of this due to the scanning phase (phase 2).

\subsubsection{Server-Side vs. Sampling Top-K} \label{sssec:topk-compare}

\begin{figure}[t!]
    \centering
    \subfloat[Runtime]{
        \includegraphics[width=0.98\columnwidth]{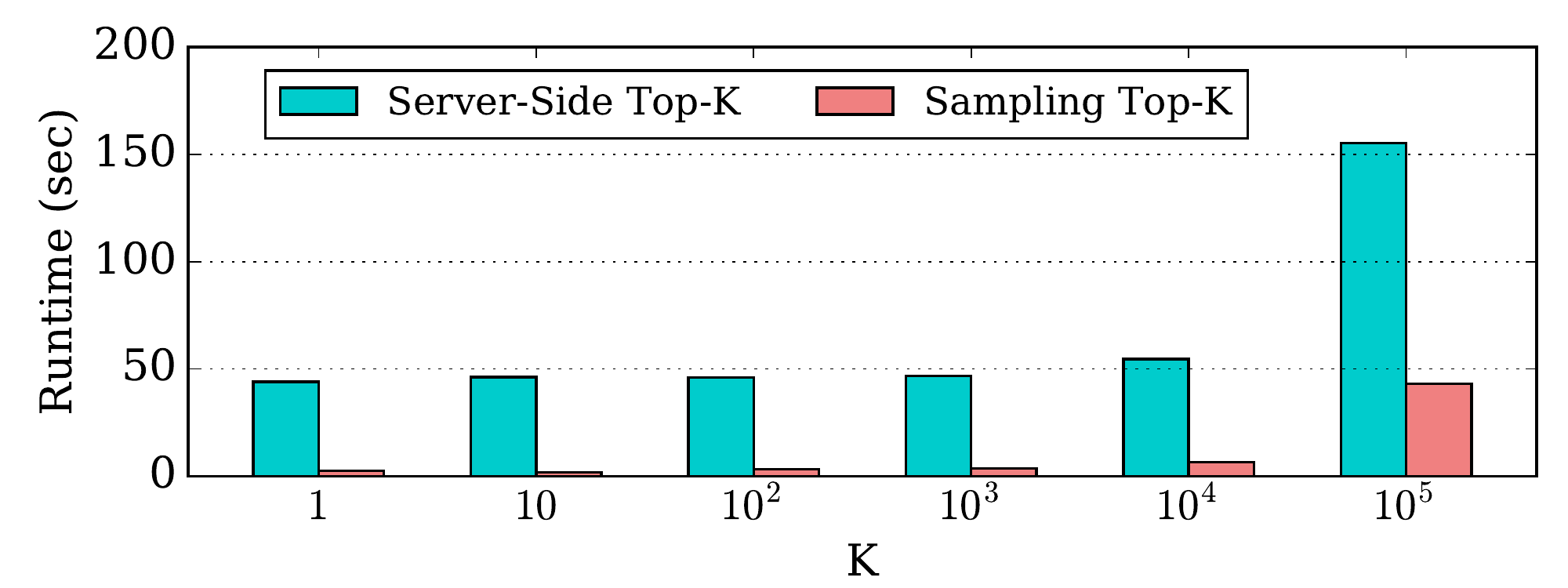}
        \label{fig:topk-k-rt}
    } \\ \vspace{-.15in}
    \subfloat[Cost]{
     \includegraphics[width=0.98\columnwidth]{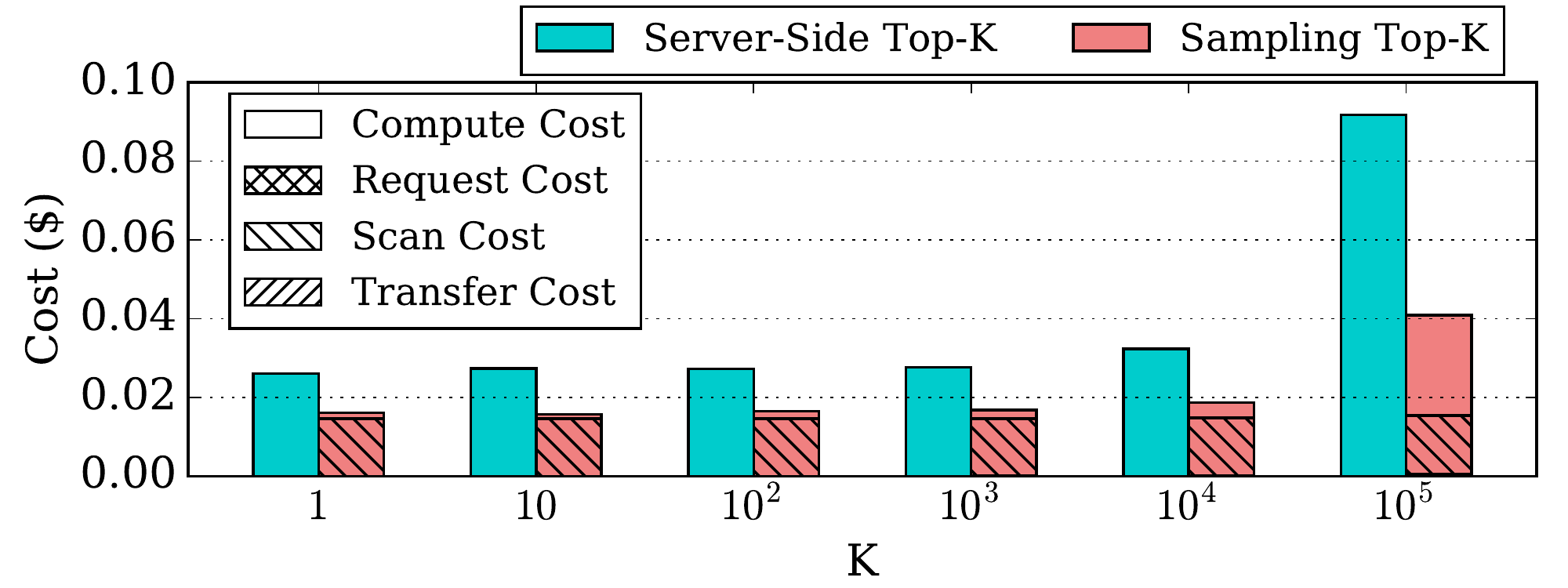}
        \label{fig:topk-k-cost}
    } \vspace{-.1in}
    \caption{
\textbf{Sensitivity to $K$} --- \normalfont{Performance and cost of server-side and sampling top-K as $K$ increases.}
    }
    \vspace{-.15in} 
    \label{fig:topk-k}
\end{figure}

We now compare the performance of the sampling-based top-K with the baseline algorithm that loads the entire table and performs top-K at the server side. $K$ is swept from $1$ to $10^5$ ($10^5$ rows comprise 0.17\% of the table). For the sampling-based algorithm, the sample size is calculated using the model in Section~\ref{ssec:topk-analysis}.

\begin{figure*}[t!]
    \centering
    \subfloat[Runtime]{
        \includegraphics[width=0.98\textwidth]{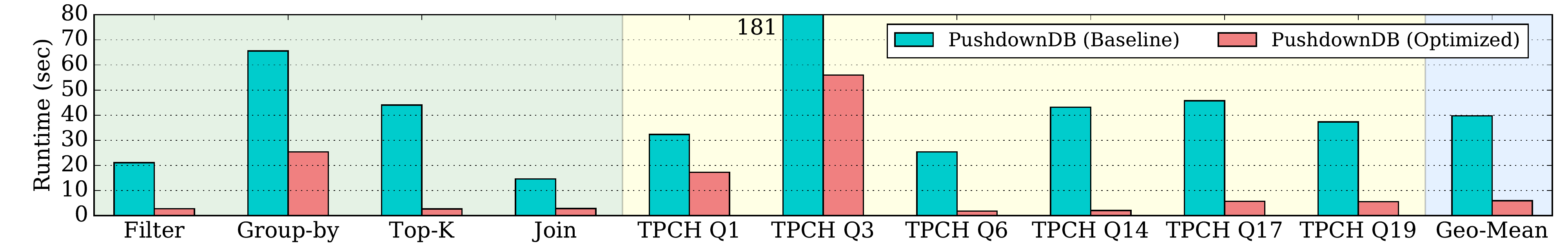}
        \label{fig:validation-rt}
    } \\ \vspace{-.15in}   
    \subfloat[Cost]{
     \includegraphics[width=0.98\textwidth]{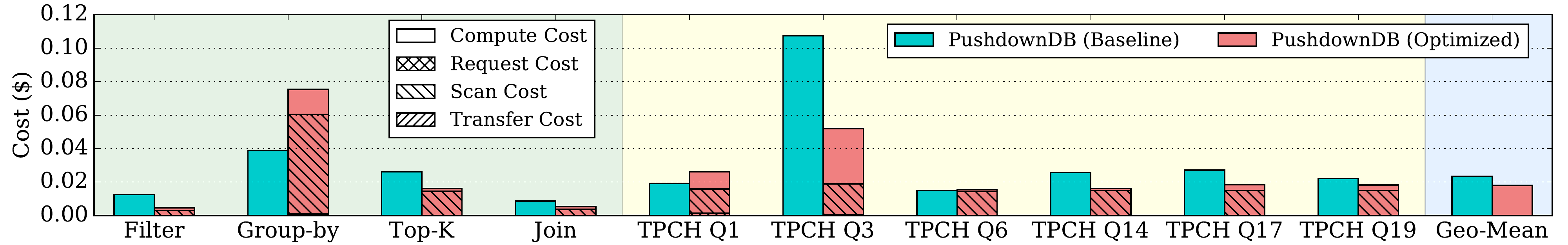}
        \label{fig:validation-cost}
    }
    \vspace{-.1in}
    \caption{
\normalfont{Performance and cost of various queries on \name.}
    } \vspace{-.2in}
    \label{fig:validation}
\end{figure*}

Figure~\ref{fig:topk-k-rt} shows that for both algorithms, runtime increases as $K$ increases. This is because a larger $K$ requires a bigger heap and also more computation at the server side. The sampling-based top-K algorithm is consistently faster than the server-side top-K due to the reduction in the amount of data loaded from S3.

In Figure~\ref{fig:topk-k-cost}, we observe that the sampling-based top-K algorithm is also consistently cheaper than server-side top-K. When $K$ is small, the majority of the cost in the sampling-based algorithm is data scanning. As $K$ increases, the data scan cost does not significantly change, but the computation cost increases due to the longer time spent obtaining the top-K using the heap.

\section{TPC-H Results} \label{sec:tpch}

In this section, we evaluate a representative query for each individual operator discussed in Sections~\ref{sec:filter} -- \ref{sec:topk}, as well as a subset of the TPC-H queries. 
Each experiment evaluates the following two configurations:

\textbf{\name (Baseline):} This is the \name
implementation described in Section~\ref{sec:testbed} but not including S3 Select features. The server loads the entire table from S3 and performs computation locally.

\textbf{\name (Optimized):} The \name that includes the optimizations discussed in this paper. 


The experiments use the 10~GB TPC-H dataset. 
The results are summarized in Figure~\ref{fig:validation}. 
From left to right, the figure shows performance and cost of individual operators (shaded in green), TPC-H queries (shaded in yellow), and geometric mean (shaded in light blue). The geo-mean cost only contains the total cost, not broken down into individual components.

As we can see, the optimizations discussed in this paper can significantly improve the performance of various types of queries. On average, the optimized \name  outperforms the baseline \name by $6.7\times$ and reduces the cost by 30\%. 
We assume a database can use various statistics of the underlying data to determine which algorithm to use for a particular query. Dynamically determining which optimization to use is orthogonal to and beyond the scope of this paper.

To validate these results, we also compared the execution time of \name to Presto, a highly optimized cloud database written in Java. We use Presto v0.205 as a performance upper bound when S3 Select is not used. On average, the runtime of baseline \name is slower than Presto by less than $2\times$, demonstrating that the code base of \name is reasonably well optimized. The optimized \name outperforms Presto by $3.4\times$.

\section{Experiments with Parquet} \label{sec:parquet}

In addition to CSV, S3 Select supports queries on the Parquet columnar data format~\cite{parquet}. In this section, we study whether Parquet offers higher performance than CSV.


\begin{figure}[t!]
    \centering
    \vspace{-.05in}
    \includegraphics[width=0.98\columnwidth]{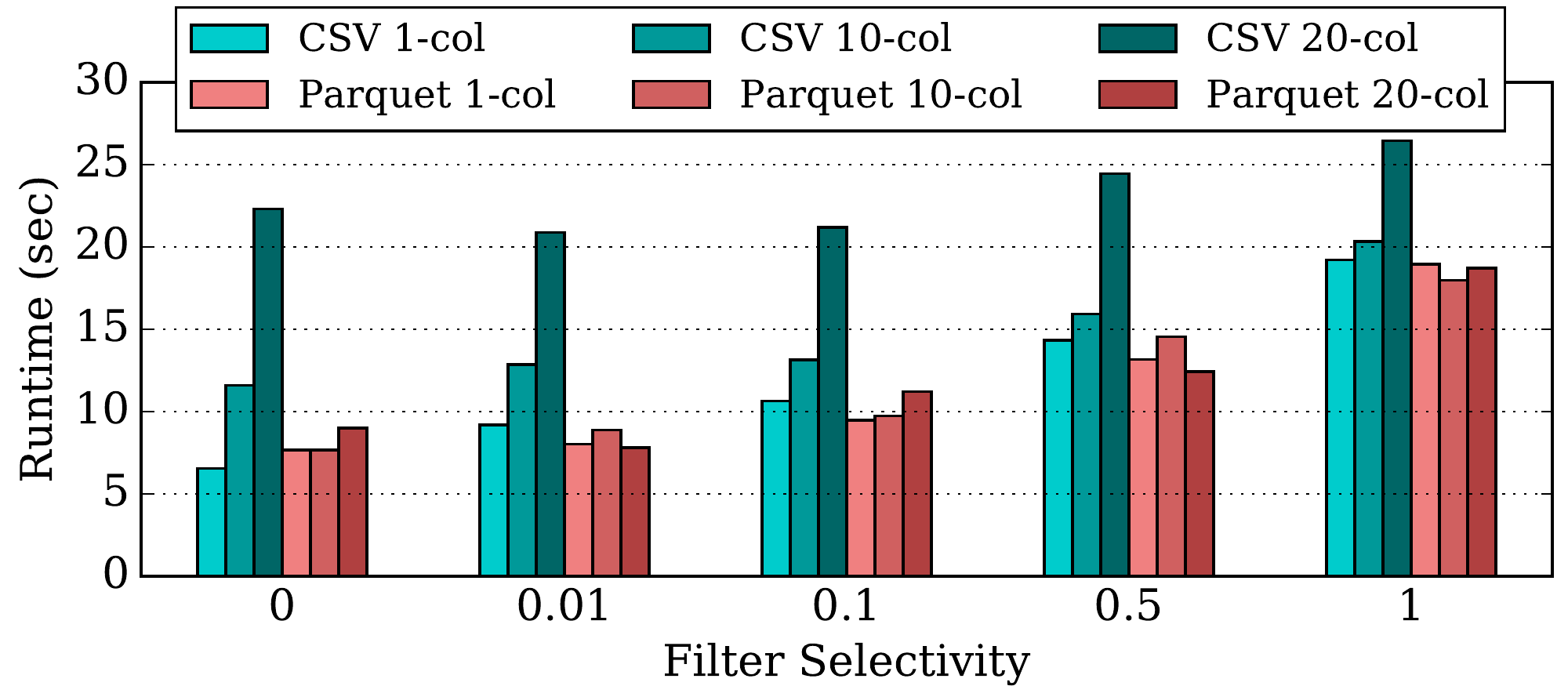}    
    \vspace{-.05in}
    \caption{
		Performance of CSV vs. Parquet}
		\vspace{-.25in}
    \label{fig:parquet}
\end{figure}

Figure~\ref{fig:parquet} shows the runtime of filter queries against both data formats. 
We implemented three tables with 1, 10, and 20 columns; each column contains 100~MB of randomly generated floating point numbers with limited precision (rounded to four decimals).
The Parquet tables use Snappy compression with a row group (i.e., logical partitioning of the data into rows) of 100~MB. 
The compressed Parquet is 70\% of its original size. 
We also tested Parquet data without compression and with different row group sizes but they lead to similar performance, and are therefore not shown here. 
The queries return a single filtered column of the table, with filtering selectivity ranging from 0 (returning no data) to 1 (returning all data).

As shown in Figure~\ref{fig:parquet}, Parquet substantially outperforms CSV in the 10 and 20 column cases, where the query requests a small fraction of columns. 
In this case, our query scans only a single column of Parquet data but has to scan the entire CSV file --- Parquet outperforms CSV due to less IO overhead on S3 Select servers. 
We also observe that the performance advantage of Parquet over CSV is more prominent when the filter is more selective --- when more data passes through, data transfer becomes the bottleneck so CSV and Parquet achieve similar performance. 
This is mainly because the current S3 Select always returns data in CSV format, even if the data is stored in Parquet format, which leads to unnecessarily large network traffic for data transfer. A potential solution to mitigate this problem is to compress transferred data. Thus, in the current S3 Select, Parquet offers a performance advantage over CSV only in extreme cases when the query touches a small fraction of columns and the data transfer over network is not a bottleneck. 

We have evaluated the same TPC-H queries as in Section~\ref{sec:tpch} on Parquet data. Although the performance numbers are not shown, Parquet on TPC-H has very limited (if any) performance advantage over CSV format. 
This is because the data accesses of TPC-H queries do not exhibit the extreme patterns as discussed above.

\section{Limitations of S3 Select} \label{sec:discussion}

So far, we have demonstrated substantial performance improvement on common database operators by leveraging S3 Select. 
In this section, we present a list of limitations of the current S3 Select features and describe our suggestions for improvement. 

\textbf{Suggestion 1: Multiple byte ranges for GET requests.}
The indexing algorithm discussed in Section~\ref{ssec:indexing} sends HTTP GET requests to S3 to load records from the table; each request asks for a specified range of bytes that are derived from an index table lookup. According the S3 API~\cite{get-object}, the current GET request to S3 supports only a single byte range. 
This means that a large number of GET requests 
have to be sent if many records are selected by a query. Excessive GET requests can hurt performance as shown in Figure~\ref{fig:filter}. 
Allowing a single GET request to contain multiple byte ranges, which is allowed by HTTP, can significantly reduce the cost of HTTP request processing in both the server and S3.

\textbf{Suggestion 2: Index inside S3.}
A more thorough solution to the indexing problem is to build the index structures entirely inside S3. 
This avoids many network messages between S3 and the server that are caused by accesses to the index data structure during an index lookup. 
S3 can handle the required logic on behalf of the server, like handling hash collisions in a hash index or traversing through the tree in a B-tree index.

\textbf{Suggestion 3: More efficient Bloom filters.}
Bloom filters can substantially improve performance of join queries, as demonstrated in Section~\ref{sec:join}. 
A Bloom filter is represented using a bit array for space efficiency. 
The current S3 Select, however, does not support bit-wise operators. 
Our current implementation of a Bloom join in S3 Select uses a string of 0’s and 1’s to represent the bit array, which is space- and computation-inefficient. 
We suggest that the next version of S3 Select should support efficient bit-wise operators to improve the efficiency of Bloom join.

\textbf{Suggestion 4: Partial group-by.}
Section~\ref{sec:hybrid-groupby} introduced our hybrid group-by algorithm and demonstrated its superior performance. Since S3 does not support group-by queries, we used the \texttt{CASE} clause to implement S3-side group-by, which is not the most efficient implementation. We suggest adding partial group-by queries to S3 to resolve this performance issue. Note that pushing an arbitrary group-by query entirely to S3 may not be the best solution, because a large number of groups can consume significant memory space and computation in S3. We consider the partial S3-side group-by as an optimization to the second phase of our current hybrid group-by.

\textbf{Suggestion 5: Computation-aware pricing.}
Across our evaluations on the optimized \name, data scan costs dominate a majority of queries. In the current S3 Select pricing model, data scanning costs a fixed amount (\$0.002/GB) regardless of what computation is being performed. 
Given that our queries typically require little computation in S3, the current pricing model may have overcharged our queries. 
We believe a fairer pricing model is needed, in which the data scan cost should depend on the workload.

\section{Related Work} \label{sec:related}


\subsection{In-Cloud Databases} \label{ssec:in-cloud-db}

Database systems are moving to the cloud environment due to cost.
Most of these in-cloud databases support storing data within S3.
Vertica~\cite{lamb2012vertica}, a traditional column-store shared nothing database, started to support S3 in its new Eon mode~\cite{vandiver2018eon}.
Snowflake~\cite{snowflake} is a software-as-a-service (SaaS) database designed specifically for the cloud environment.
Many open-source in-cloud databases have been developed and widely adopted, examples including Presto~\cite{presto}, Hive~\cite{hive}, and Spark SQL~\cite{sparksql}.
Furthermore, AWS offers a few proprietary database systems in the cloud: Athena~\cite{athena}, Aurora~\cite{aurora}, and Redshift~\cite{spectrum}.

Among the systems mentioned above, Presto, Spark, and Hive support S3 Select in Amazon Elastic MapReduce (EMR) in limited form. For example, Presto supports pushing predicates to S3 but does not support data types like timestamp, real, or double.
Furthermore, these systems currently support only simple filtering operations but not complex ones like join, group-by, or top-K, which are what \name focuses on.

The Spectrum feature of Redshift offloads some query processing on data stored in S3 to the ``Redshift Spectrum Layer'' such that more parallelism can be exploited beyond the capability of the cluster created by the user.
The ideas discussed in this paper can be applied to the Redshift Spectrum setting to improve performance of complex database operators.

\subsection{Database Machines} \label{ssec:db-machine}

A line of research on database machines emerged in the 1970s and stayed active for more than 10 years.
These systems contain processors or special hardware to accelerate database accesses, by applying the principle of pushing computation to where the data resides.

The Intelligent Database Machines (IDM)~\cite{ubell1985intelligent} from Britton Lee separated the functionality of host computers and the database machine which sits closer to the disks.
Much of a DBMS functionality can be performed on the database machine, thereby freeing the host computers to perform other tasks.
Grace ~\cite{fushimi1986overview} is a parallel database machine that contains multiple processors connected to multiple disk modules.
Each disk module contains a filter processor that can perform selection using predicates and projection to reduce the amount of data transfer as well as computation in the main processors.

More recently, in the 2000s, IBM Netezza data warehouse appliances~\cite{netezza} used FPGA-enabled near-storage processors (FAST engines) to support data compression, projection, and row selection.
In Oracle's Exadata~\cite{exadata} database engines, the storage unit (Exadata Cell) can support predicate filtering, column filtering, Bloom join, encryption, and indexing among other functionalities.

\subsection{Near-Data Processing (NDP)}
\label{ssec:ndp-db}

Near-data processing has recently attracted much research interest in the computer architecture community~\cite{balasubramonian2014near}. Techniques have been proposed for memory and storage devices in various part of the system. Although the techniques in this paper were proposed assuming a cloud storage setting, many of them can be applied to the following other settings as well.

Processing-in-Memory (PIM)~\cite{ghose2018enabling} exploits computation near or inside DRAM devices to reduce data transfer between CPU and main memory, which is a bottleneck in modern processors. Recent development in 3D-stacked DRAM implements logic at the bottom layer of the memory chip~\cite{hmc}, supporting in-memory processing with lower energy and cost.

While smart disks have been studied in the early 2000s~\cite{riedel2001active, keeton1998case}, they have not seen wide adoption due to the limitations of the technology.
The development of FPGAs and SSDs in recent years has made near storage computing more practical.
Recent studies have proposed to push computation to both near-storage FPGAs~\cite{woods2014ibex, gao2016hrl} and the processor within an SSD device~\cite{gu2016biscuit, koo2017summarizer, do2013query}. Most of these systems only focused on simple operators like filter or projection, but did not study the effect of more complex operators as we do in \name.

Hybrid shipping techniques execute some query operators at the client side, where the query is invoked, and some at the server side, where data is stored~\cite{franklin1996performance}. However, near-storage computing services as S3 do not support complex operators such as joins. Hybrid shipping does not consider how to push down only some of the steps involved in the implementation of a single operator, which is what \name addresses.

\section{Conclusion} \label{sec:conclusion}

This paper presents \name, a data analytics engine that accelerates common database operators by performing computation in S3 via S3 Select. \name reduces both runtime and cost for a wide range of operators, including filter, project, join, group-by, and top-K. Using S3 Select, \name improves the average performance of a subset of the TPC-H queries by $6.7\times$ and reduces cost by 30\%. 

\small
\bibliographystyle{IEEEtran}
\bibliography{main}

\begin{thebibliography}{10}
\providecommand{\url}[1]{#1}
\csname url@samestyle\endcsname
\providecommand{\newblock}{\relax}
\providecommand{\bibinfo}[2]{#2}
\providecommand{\BIBentrySTDinterwordspacing}{\spaceskip=0pt\relax}
\providecommand{\BIBentryALTinterwordstretchfactor}{4}
\providecommand{\BIBentryALTinterwordspacing}{\spaceskip=\fontdimen2\font plus
\BIBentryALTinterwordstretchfactor\fontdimen3\font minus
  \fontdimen4\font\relax}
\providecommand{\BIBforeignlanguage}[2]{{%
\expandafter\ifx\csname l@#1\endcsname\relax
\typeout{** WARNING: IEEEtran.bst: No hyphenation pattern has been}%
\typeout{** loaded for the language `#1'. Using the pattern for}%
\typeout{** the default language instead.}%
\else
\language=\csname l@#1\endcsname
\fi
#2}}
\providecommand{\BIBdecl}{\relax}
\BIBdecl

\bibitem{presto}
``{Presto},'' \url{https://prestodb.io}, 2018.

\bibitem{snowflake}
B.~Dageville, T.~Cruanes, M.~Zukowski, V.~Antonov, A.~Avanes, J.~Bock,
  J.~Claybaugh, D.~Engovatov, M.~Hentschel, J.~Huang \emph{et~al.}, ``{The
  Snowflake Elastic Data Warehouse},'' in \emph{SIGMOD}, 2016.

\bibitem{spectrum}
``{Amazon Redshift},'' \url{https://aws.amazon.com/redshift/}, 2018.

\bibitem{clouddb}
J.~Tan, T.~Ghanem, M.~Perron, X.~Yu, M.~Stonebraker, D.~DeWitt, M.~Serafini,
  A.~Aboulnaga, and T.~Kraska, ``{Choosing A Cloud DBMS: Architectures and
  Tradeoffs},'' in \emph{VLDB}, 2019.

\bibitem{gupta2015amazon}
A.~Gupta, D.~Agarwal, D.~Tan, J.~Kulesza, R.~Pathak, S.~Stefani, and
  V.~Srinivasan, ``{Amazon Redshift and the Case for Simpler Data
  Warehouses},'' in \emph{SIGMOD}, 2015.

\bibitem{hagmann1986performance}
R.~B. Hagmann and D.~Ferrari, ``Performance analysis of several back-end
  database architectures,'' \emph{ACM Transactions on Database Systems (TODS)},
  vol.~11, no.~1, pp. 1--26, 1986.

\bibitem{ubell1985intelligent}
M.~Ubell, ``{The Intelligent Database Machine (IDM)},'' in \emph{Query
  processing in database systems}.\hskip 1em plus 0.5em minus 0.4em\relax
  Springer, 1985, pp. 237--247.

\bibitem{exadata}
R.~Weiss, ``{A Technical Overview of the Oracle Exadata Database Machine and
  Exadata Storage Server},'' \emph{Oracle White Paper. Oracle Corporation,
  Redwood Shores}, 2012.

\bibitem{netezza}
P.~Francisco, ``{The Netezza Data Appliance Architecture},'' 2011.

\bibitem{s3select}
R.~Hunt, ``{S3 Select and Glacier Select – Retrieving Subsets of Objects},''
  \url{https://aws.amazon.com/blogs/aws/s3-glacier-select/}, 2018.

\bibitem{s3}
``{Amazon S3},'' \url{https://aws.amazon.com/s3/}, 2018.

\bibitem{hive}
A.~Thusoo, J.~S. Sarma, N.~Jain, Z.~Shao, P.~Chakka, N.~Zhang, S.~Antony,
  H.~Liu, and R.~Murthy, ``{Hive --- A Petabyte Scale Data Warehouse Using
  Hadoop},'' in \emph{ICDE}, 2010.

\bibitem{sparksql}
M.~Armbrust, R.~S. Xin, C.~Lian, Y.~Huai, D.~Liu, J.~K. Bradley, X.~Meng,
  T.~Kaftan, M.~J. Franklin, A.~Ghodsi \emph{et~al.}, ``{Spark SQL: Relational
  Data Processing in Spark},'' in \emph{SIGMOD}, 2015.

\bibitem{parquet}
``{Apache Parquet},'' \url{https://parquet.apache.org}, 2016.

\bibitem{mckinney2011pandas}
W.~McKinney, ``{pandas: a Foundational Python Library for Data Analysis and
  Statistics},'' \emph{Python for High Performance and Scientific Computing},
  pp. 1--9, 2011.

\bibitem{bloom1970space}
B.~H. Bloom, ``{Space/Time Trade-offs in Hash Coding with Allowable Errors},''
  \emph{Communications of the ACM}, vol.~13, no.~7, pp. 422--426, 1970.

\bibitem{universal_hashing}
J.~L. Carter and M.~N. Wegman, ``Universal classes of hash functions,''
  \emph{Journal of Computer and System Sciences}, vol.~18, no.~2, pp. 143--154,
  1979.

\bibitem{scalable_bloom_filters}
P.~Almeida, C.~Baquero, N.~Preguica, and D.~Hutchison, ``Scalable bloom
  filters,'' \emph{Information Processing Letters}, vol. 101, no.~6, pp.
  255--261, 2007.

\bibitem{gray1994quickly}
J.~Gray, P.~Sundaresan, S.~Englert, K.~Baclawski, and P.~J. Weinberger,
  ``{Quickly Generating Billion-Record Synthetic Databases},'' in \emph{Acm
  Sigmod Record}, vol.~23, no.~2, 1994, pp. 243--252.

\bibitem{get-object}
``{Amazon Simple Storage Service, GET Object},''
  \url{https://docs.aws.amazon.com/AmazonS3/latest/API/RESTObjectGET.html},
  2006.

\bibitem{lamb2012vertica}
A.~Lamb, M.~Fuller, R.~Varadarajan, N.~Tran, B.~Vandiver, L.~Doshi, and
  C.~Bear, ``{The Vertica Analytic Database: C-Store 7 Years Later},''
  \emph{VLDB}, 2012.

\bibitem{vandiver2018eon}
B.~Vandiver, S.~Prasad, P.~Rana, E.~Zik, A.~Saeidi, P.~Parimal, S.~Pantela, and
  J.~Dave, ``{Eon Mode: Bringing the Vertica Columnar Database to the Cloud},''
  in \emph{SIGMOD}, 2018.

\bibitem{athena}
``{Amazon Athena --- Serverless Interactive Query Service},''
  \url{https://aws.amazon.com/athena/}, 2018.

\bibitem{aurora}
A.~Verbitski, A.~Gupta, D.~Saha, M.~Brahmadesam, K.~Gupta, R.~Mittal,
  S.~Krishnamurthy, S.~Maurice, T.~Kharatishvili, and X.~Bao, ``{Amazon Aurora:
  Design Considerations for High Throughput Cloud-Native Relational
  Databases},'' in \emph{SIGMOD}, 2017.

\bibitem{fushimi1986overview}
S.~Fushimi, M.~Kitsuregawa, and H.~Tanaka, ``{An Overview of The System
  Software of A Parallel Relational Database Machine GRACE},'' in \emph{VLDB},
  1986.

\bibitem{balasubramonian2014near}
R.~Balasubramonian, J.~Chang, T.~Manning, J.~H. Moreno, R.~Murphy, R.~Nair, and
  S.~Swanson, ``{Near-Data Processing: Insights from a MICRO-46 Workshop},''
  \emph{IEEE Micro}, 2014.

\bibitem{ghose2018enabling}
S.~Ghose, K.~Hsieh, A.~Boroumand, R.~Ausavarungnirun, and O.~Mutlu, ``{Enabling
  the Adoption of Processing-in-Memory: Challenges, Mechanisms, Future Research
  Directions},'' \emph{arXiv preprint arXiv:1802.00320}, 2018.

\bibitem{hmc}
HybridMemoryCubeConsortium, ``{HMCSpecification2.1},'' 2014.

\bibitem{riedel2001active}
E.~Riedel, C.~Faloutsos, G.~A. Gibson, and D.~Nagle, ``Active disks for
  large-scale data processing,'' \emph{Computer}, vol.~34, no.~6, pp. 68--74,
  2001.

\bibitem{keeton1998case}
K.~Keeton, D.~A. Patterson, and J.~M. Hellerstein, ``A case for intelligent
  disks (idisks),'' \emph{ACM SIGMOD Record}, vol.~27, no.~3, pp. 42--52, 1998.

\bibitem{woods2014ibex}
L.~Woods, Z.~Istv{\'a}n, and G.~Alonso, ``{Ibex: an Intelligent Storage Engine
  with Support for Advanced SQL Offloading},'' \emph{VLDB}, 2014.

\bibitem{gao2016hrl}
M.~Gao and C.~Kozyrakis, ``{HRL: Efficient and Flexible Reconfigurable Logic
  for Near-Data Processing},'' in \emph{HPCA}, 2016.

\bibitem{gu2016biscuit}
B.~Gu, A.~S. Yoon, D.-H. Bae, I.~Jo, J.~Lee, J.~Yoon, J.-U. Kang, M.~Kwon,
  C.~Yoon, S.~Cho \emph{et~al.}, ``Biscuit: A framework for near-data
  processing of big data workloads,'' in \emph{ISCA}, 2016.

\bibitem{koo2017summarizer}
G.~Koo, K.~K. Matam, H.~Narra, J.~Li, H.-W. Tseng, S.~Swanson, M.~Annavaram
  \emph{et~al.}, ``Summarizer: trading communication with computing near
  storage,'' in \emph{MICRO}, 2017.

\bibitem{do2013query}
J.~Do, Y.-S. Kee, J.~M. Patel, C.~Park, K.~Park, and D.~J. DeWitt, ``{Query
  Processing on Smart SSDs: Opportunities and Challenges},'' in \emph{SIGMOD},
  2013.

\bibitem{franklin1996performance}
M.~J. Franklin, B.~T. J{\'o}nsson, and D.~Kossmann, ``Performance tradeoffs for
  client-server query processing,'' in \emph{ACM SIGMOD Record}, vol.~25,
  no.~2.\hskip 1em plus 0.5em minus 0.4em\relax ACM, 1996, pp. 149--160.

\end{thebibliography}

\end{document}